\documentclass[journal]{IEEEtran}

\newcommand{\beq}{\begin{equation}}
\newcommand{\eeq}{\end{equation}}
\newcommand{\bsq}{\begin{subequations}}
\newcommand{\esq}{\end{subequations}}
\newcommand{\bq}{\begin{eqnarray}}
\newcommand{\eq}{\end{eqnarray}}
\newcommand{\bqn}{\begin{eqnarray*}}
\newcommand{\eqn}{\end{eqnarray*}}
\usepackage{bbm}
\usepackage[pdftex]{graphicx}
\graphicspath{{Figs/}}
\usepackage{enumerate}
\usepackage{enumitem} 

\usepackage{mathptmx}       
\DeclareMathAlphabet{\mathcal}{OMS}{cmsy}{m}{n}
\usepackage{helvet}         
\usepackage{courier}        
\usepackage{type1cm}        
\usepackage{makeidx}         
\usepackage{graphicx}        
\usepackage{url} 
\usepackage{multicol}        
\usepackage[bottom]{footmisc}
\usepackage{ifthen}
\usepackage{subfigure}
\usepackage[usenames,dvipsnames]{color}

\makeindex             

\usepackage{graphics} 
\usepackage{graphicx} 
\usepackage{epsfig} 
\usepackage{epstopdf}
\usepackage{mathptmx} 
\usepackage{times} 
\usepackage[cmex10]{amsmath}

\usepackage{mathtools}  
\usepackage{amssymb}  
\usepackage{amsthm}

\usepackage{amsfonts}
\usepackage{cite}
\usepackage{verbatim}
\usepackage{comment}
\usepackage{algorithmic}
\usepackage{multirow}
\usepackage{cite}
\usepackage{stfloats}
\usepackage{supertabular}
\usepackage{longtable}

\DeclareGraphicsExtensions{.pdf,.png,.jpg,.eps,.ps}
\renewcommand{\arraystretch}{1.2}
\usepackage{arydshln}
\usepackage{multicol}
\usepackage{colortbl}
\theoremstyle{definition}
\newtheorem{theorem}{Theorem}

\theoremstyle{definition}

\ifCLASSINFOpdf
\else
\fi

\usepackage{comment}
\usepackage{ifthen}
\newboolean{showcomments}
\setboolean{showcomments}{true}
\newcommand{\ychen}[1]{\ifthenelse{\boolean{showcomments}}
        { \textcolor{red}{YC: #1}}}

\usepackage{amsmath}
\allowdisplaybreaks[4]

\usepackage[ruled]{algorithm2e}

\begin{document}

%
\title{Sizing Grid-Connected Wind Power Generation and Energy Storage with Wake Effect and Endogenous Uncertainty: A Distributionally Robust Method}

%

\author{Rui Xie,~
       Wei Wei,~\IEEEmembership{Senior Member,~IEEE},~and Yue Chen,~\IEEEmembership{Member,~IEEE}
\thanks{R. Xie and Y. Chen are with the Department of Mechanical and Automation Engineering, the Chinese University of Hong Kong, Hong Kong SAR. (email: ruixie@cuhk.edu.hk; yuechen@mae.cuhk.edu.hk)}
\thanks{W. Wei is with the State Key Laboratory of Power Systems, Department of Electrical Engineering, Tsinghua University, Beijing 100084, China. (email: wei-wei04@mails.tsinghua.edu.cn)}
}

%
%

\markboth{CSEE Journal of Power and Energy Systems,~Vol.~XX, No.~X, Feb.~2019}%
{Xie \MakeLowercase{\textit{et al.}}: Sizing Grid-Connected Wind Power Generation and Energy Storage with Wake Effect and Endogenous Uncertainty: A Distributionally Robust Method}
%



\maketitle

\begin{abstract}
Wind power, as a green energy resource, is growing rapidly worldwide, along with energy storage systems (ESSs) to mitigate its volatility. Sizing of wind power generation and ESSs has become an important problem to be addressed. Wake effect in a wind farm can cause wind speed deficits and a drop in downstream wind turbine power generation, which however was rarely considered in the sizing problem in power systems. In this paper, a bi-objective distributionally robust optimization (DRO) model is proposed to determine the capacities of wind power generation and ESSs considering the wake effect. An ambiguity set based on Wasserstein metric is established to characterize the wind power and demand uncertainties. In particular, wind power uncertainty is affected by the wind power generation capacity which is determined in the first stage. Thus, the proposed model is a DRO problem with endogenous uncertainty (or decision-dependent uncertainty). To solve the proposed model, a stochastic programming approximation method based on minimum Lipschitz constants is developed to turn the DRO model into a linear program. Then, an iterative algorithm is built, embedded with methods for evaluating the minimum Lipschitz constants. Case studies demonstrate the necessity of considering wake effect and the effectiveness of the proposed method.
\end{abstract}


%
\IEEEpeerreviewmaketitle

\section{Introduction}

\IEEEPARstart{T}{he} large-scale integration of renewable energy is one of the worldwide efforts towards carbon neutrality \cite{wei2022bi}. Wind power, as one of the leading clean energy resources, has developed rapidly in recent years. However, wind resources are affected by weather conditions, leading to highly volatile and stochastic wind power, which threatens the power system operation. Energy storage systems (ESSs) that can shift energy over time are installed to better accommodate uncertain renewable generations. Therefore, the sizing of wind power generation and ESSs is an important problem to be addressed.

A vast literature has been devoted to the generation and ESS siting and sizing problems in renewable-integrated power systems. These studies can be divided into three categories based on stochastic programming (SP), robust optimization (RO), and distributionally robust optimization (DRO), respectively. The SP methods assume known probability distributions of the uncertainties, which are represented either by parametrized distributions or a set of scenarios. The sizing problem of generation and storage expansion \cite{canas2020generation}, transmission and ESS capacities \cite{bhattacharjee2020benefits}, and hybrid ESS systems \cite{wan2021probabilistic} were studied using the SP method. In \cite{yin2022coordinated}, the reduction of wind power prediction error caused by the aggregation effect was modeled by decision-dependent Gaussian distributions and further represented by generated scenarios.
However, in practice, it is difficult to obtain the accurate uncertainty distribution in the planning stage, and an inexact empirical distribution may lead to infeasible or suboptimal results. To overcome this shortcoming, RO has been adopted, which optimizes the performance under the worst-case scenario within an uncertainty set. A tri-level robust model was proposed for ESS planning \cite{liu2020accommodating}, where the uncertainties of wind power investment and coal-fired unit retirement were considered. 
Reference \cite{yan2022two} jointly considered the short-term fluctuations and the long-term uncertainties of load and renewable growths in the ESS planning problem. However, as the worst case rarely happens, the RO method may lead to over-conservative sizing results. 

DRO technique is in-between SP and RO methods, which optimizes decisions with respect to the worst-case distribution in a predefined ambiguity set. It can achieve a good tradeoff between optimality and robustness. 
DRO method has appeared in power system operation problems. Moments were used in \cite{cao2021day} to construct the ambiguity set for distributionally robust energy management of energy hubs. 
The uncertain market price was modeled by ambiguity sets based on $\phi$-divergence in the self-scheduling of compressed air energy storage in \cite{li2021risk}.
DRO method has also been adopted in power system sizing problems. 
Aiming at optimal generation expansion, a DRO model with risk constraints was put forward in \cite{pourahmadi2021distributionally} with an ambiguity set based on moments and unimodality. However, large numbers of data are needed to accurately estimate moments and different distributions may have the same moments. A DRO siting and sizing method for ESS was proposed in \cite{le2021distributionally} based on $\phi$-divergence, where the ESS lifespan model was considered. However, $\phi$-divergence is only well-defined for discrete distributions supported on the range of the sample data. On the contrary, Wasserstein distance is able to measure the difference between general distributions. A distributionally robust chance-constrained model was proposed for clustered generation expansion in \cite{chen2022wasserstein} with a Wasserstein-metric-based ambiguity set. However, none of the studies above considered the wake effect in a wind farm.

The wake effect in a wind farm refers to the decrease in gross energy production due to the changes in wind speed caused by the impact of wind turbines (WTs) on each other.
To be specific, the wake from the upstream WTs decreases the wind's kinetic energy and lowers the wind speed at the downstream WTs. Studies on wind farms have been aware of the wake effect and considered it in problems such as reliability evaluation \cite{ren2018reliability}, dynamic equivalent model \cite{gupta2022multi}, control strategy \cite{lyu2019novel}, maintenance scheduling \cite{yin2020decision}, etc. As the wake effect can reduce the power output by about 20\% \cite{gao2016optimization}, it is nonnegligible in wind farm planning. Currently, the wake effect is mainly considered in the layout design of wind farms. For instance, the wind farm layout was optimized in \cite{ma2021bi} considering the wake effect and the temporal correlation of wind speed. Another layout optimization approach was proposed in \cite{bai2022wind} based on an adaptive evolutionary algorithm with Monte Carlo tree search reinforcement learning. The comprehensive problem of wind farm area, shape, and layout decisions was studied in \cite{cazzaro2022multi} and the method was applied to an offshore wind farm in the UK. The optimal ESS allocation inside the wind farm was investigated in \cite{xiong2021optimal} to better support frequency. However, the aforementioned studies did not consider the impact of wind power generation on power system operation. The SP method proposed in \cite{sadeghian2020clustering} planned wind farms in distribution systems, where the wake effect was integrated via simulations. The optimal solution was searched in an enumeration manner with an acceleration constraint, which may be computationally expensive for bulk power systems with a large number of WTs.


This paper proposes a distributionally robust optimization method with a decision-dependent ambiguity set for sizing wind power generation and energy storage considering the wake effect. The contributions are:

(1) \textbf{Sizing Model}. A distributionally robust optimization model is developed for the sizing of wind power generation and ESSs in power systems. The worst-case expected fuel cost in normal operation conditions is minimized while the worst-case expected load shedding in extreme conditions is upper bounded in the constraint. Distinct from the traditional models, the wake effect in a wind farm is considered. This makes the sizing model more practical. In particular, simulations are carried out to evaluate the wake effect. Based on this, a piecewise linear model showing how the available wind power generation changes with wind conditions and the capacity is developed. The model is then used for modeling wind power uncertainty via Wasserstein-metric-based ambiguity sets. Due to the wake effect, the size of the ambiguity set depends on the first-stage sizing decisions, and is thus decision-dependent. Therefore, the proposed model is a DRO problem with both endogenous/decision-dependent uncertainty (DDU) and exogenous/decision-independent uncertainty (DIU).

(2) \textbf{Solution Method}. To solve the proposed model, we first develop an SP approximation of it with the help of minimum Lipschitz constants. Then, an iterative algorithm embedded with methods for calculating the upper bounds of the minimum Lipschitz constants is established to solve the obtained SP problem. The proposed algorithm is novel in that it provides a systematic method for solving the Wasserstein-metric-based DRO model and can deal with DDUs that are seldom considered in previous work.


The rest of this paper is organized as follows. Section \ref{sec-2} establishes a linear approximation of available wind power function considering the wake effect. A distributionally robust wind-storage sizing model that takes into account the wake effect is proposed in Section \ref{sec-3} with solution method developed in Section \ref{sec-4}. Case studies are reported in Section \ref{sec-5}. Finally, Section \ref{sec-6} concludes the paper. 

\section{Linearized Available Wind Power Function Considering Wake Effect}
\label{sec-2}

In this section, we characterize the available wind power in a wind farm considering the wake effect. First, the Jensen wake model is adopted to model the wake effect of a single WT. Simulations are carried out to calculate the available wind power in a wind farm under different total WT capacities and real-time wind conditions, whose relationship is a nonlinear function. Then, the nonlinear function is further linearized for computational tractability.

\subsection{Available Wind Power By Simulation}

Jensen wake model is a widely used analytical wake model \cite{gonzalez2014review}. Suppose the wind direction is parallel to the line between two WTs, then according to the Jensen wake model, the wind speed at the downstream WT can be calculated as follows \cite{feijoo2000modeling}:
\begin{align}
\label{eq:Jensen}
    v_2^W & = v_1^W \left( 1 - \left(1 - \sqrt{1 - C^T}\right) \left(\frac{D^R}{D^R + 2K^W d^W}\right)^2 \right), \\
    K^W & = 0.5 / \ln{(H^H/Z^R)},
\end{align}
where $v_1^W$ and $v_2^W$ are the wind speed values at the upstream and downstream WTs, respectively; $d^W$ is the distance between the two WTs; the thrust factor $C^T$, the turbine rotor diameter $D^R$, the hub height $H^H$, and the roughness length $Z^R$ are parameters; $K^W$ is the wake decay constant. As \eqref{eq:Jensen} shows, the closer the two WTs to each other, the stronger the wake and the wind speed deficit.

\begin{figure}[t]
\centering
\includegraphics[width=0.7\columnwidth]{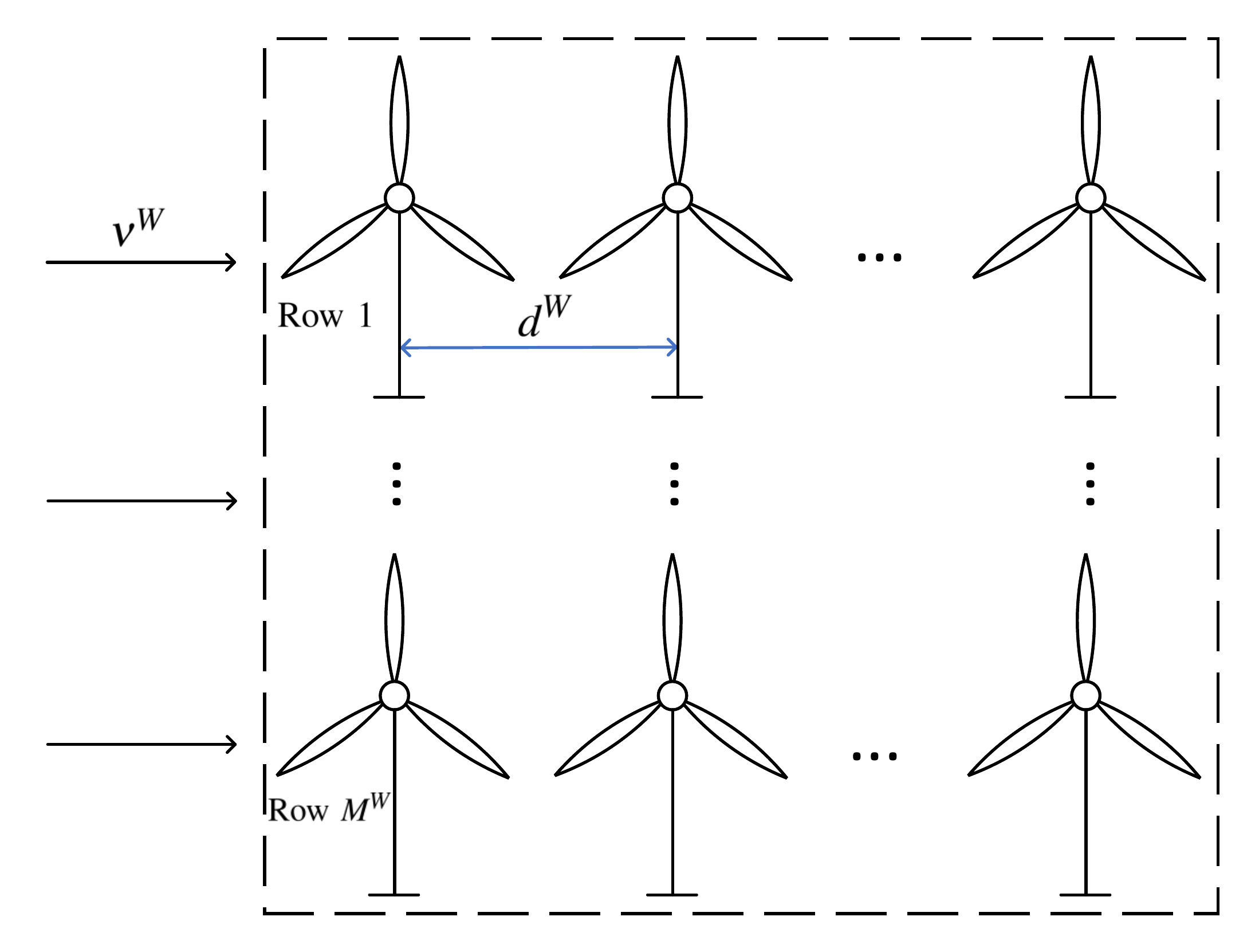}
\caption{Wind farm layout as an example.}
\label{fig:WindFarm}
\end{figure}

For simplicity, we consider the wind farm layout in Fig. \ref{fig:WindFarm} as an example: The wind farm will be built in a given rectangular area; all WTs have the same types and parameters and are placed evenly; the wind is blowing parallel to the rows of WTs; the distance between adjacent rows is fixed and large enough so that the wake effect between rows can be neglected. A similar assumption can be found in \cite{feijoo2000modeling}. Nonetheless, other wind farm layouts as well as wake models can be used for simulation and the subsequent process from Section \ref{sec:2-B} to Section \ref{sec-4} can still be carried out. Suppose the length of a row is $R^P$ and the number of WTs in a row is $n^W > 1$, then the distance between adjacent WTs is $d^W = R^P/(n^W-1)$. The wind speed at the $i$-th WT is
\begin{align}
    v_i^W & = v_1^W \left( 1 - \left(1 - \sqrt{1 - C^T}\right) \left(\frac{D^R}{D^R + 2K^W d^W}\right)^2 \right)^{i-1}.
\end{align}

The available wind power of one WT can be calculated using the following function \cite{villanueva2011probabilistic}.
\begin{align}
\label{eq:pw}
    p^W(v) := \left\{
    \begin{array}{ll}
    0, & ~\mbox{if}~ v \leq V^C~\mbox{or}~v > V^F, \\
    P^R \frac{v^2 - (V^C)^2}{(V^R)^2 - (V^C)^2}, & ~\mbox{if}~ V^C < v \leq V^R, \\
    P^R, & ~\mbox{if}~ V^R < v \leq V^F, 
    \end{array}
    \right.
\end{align}
where $p^W$ is the available wind power and $v$ is the wind speed at this WT; the cut-in speed $V^C$, cut-out speed $V^F$, rated speed $V^R$, and rated power $P^R$ are parameters. Sum up the available power of all WTs, then we have the available wind power of the wind farm $f^W := M^W\sum_{i=1}^{n^W} p^W(v_i^W)$, where $M^W$ is the number of rows. Given $n^W$ and the initial wind speed $\nu^W := v_1^W$ at the wind farm, the total capacity of WTs $x^W = P^R M^W n^W$ and the total available wind power $f^W$ can be uniquely determined. Hence, we can characterize the change of available wind power with $x^W$ and $\nu^W$ by simulations. 

\begin{figure}[t]
\centering
\includegraphics[width=1.0\columnwidth]{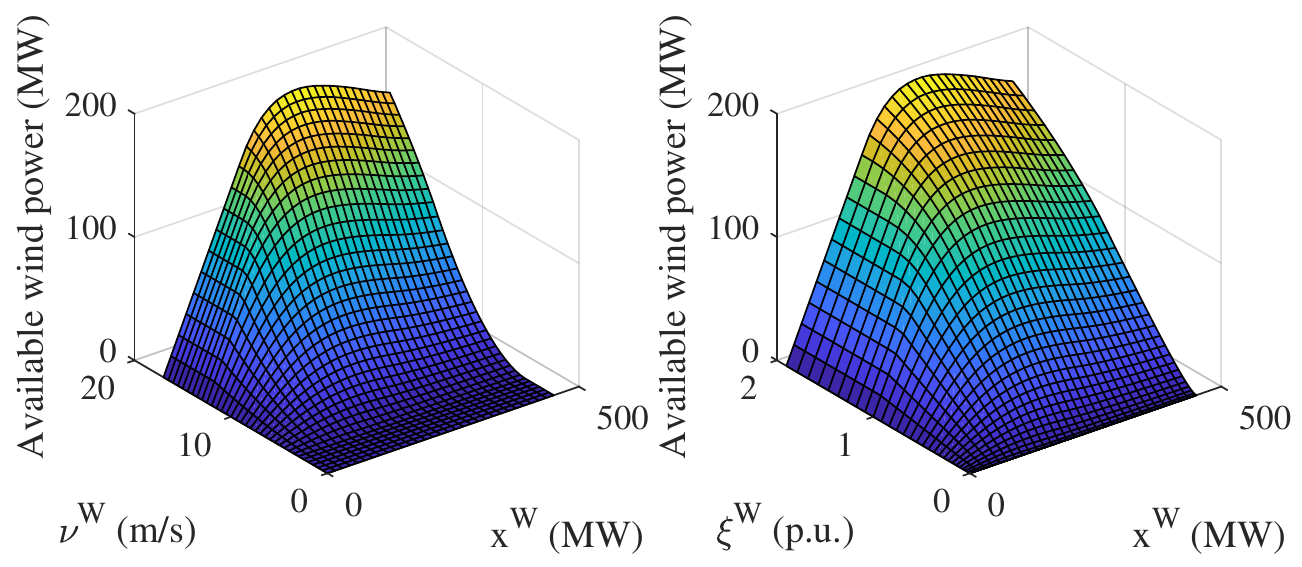}
\caption{The change of available wind power $f^W$ with the total capacity $x^W$ and initial wind speed $\nu^W$ / auxiliary variable $\xi^W$.}
\label{fig:WakeSpeedXi}
\end{figure}

To be specific, let $\bar{n}^W$ and $\bar{\nu}^W$ denote the upper bounds of $n^W$ and $\nu^W$, respectively. Use $N^V+1$ wind speed values for discretization. For $n^W = 1,2,\cdots,\bar{n}^W$ and $\nu^W = 0,\bar{\nu}^W/N^V,2\bar{\nu}^W/N^V,\dots,\bar{\nu}^W$, calculate the corresponding $f^W$ and $x^W$ and get tuples $(x_l^W,\nu_l^W,f_l^W)$ for $l = 1,2,\dots,\bar{n}^W(N^V+1)$. The change of $f^W$ with $x^W$ and $v^W$ is shown in Fig. \ref{fig:WakeSpeedXi} (left). When $\nu^W$ is fixed, as $x^W$ varies from 0 to $\bar{x}^W := P^R M^W \bar{n}^W$, $f^W$ first increases proportionally, but then its growth slows down; $f^W$ even decreases when $x^W$ is too large because of the severe wake effect. This shows the necessity of considering the wake effect in the sizing problem.

\subsection{Linearized Available Wind Power Function}
\label{sec:2-B}
Considering that the relationship between $f^W$ and $x^W, v^W$ is highly nonlinear, to facilitate the computation, we establish a piecewise linear approximation for the available wind power function.
First, to eliminate the nonlinear term in \eqref{eq:pw}, we introduce an auxiliary variable $\xi^W$ as follows.
\begin{align}
\label{eq:xi}
    \xi^W(\nu^W) := \left\{
    \begin{array}{ll}
    0, & ~\mbox{if}~ \nu^W \leq V^C, \\
    \frac{(\nu^W)^2 - (V^C)^2}{(V^R)^2 - (V^C)^2}, & ~\mbox{if}~ \nu^W > V^C.
    \end{array}
    \right.
\end{align}
Then, the available wind power $f^W$ can be viewed as a function of $x^W$ and $\xi^W$, which is depicted in Fig. \ref{fig:WakeSpeedXi} (right). The actual wind power output can vary within 0 and $f^W$, so the feasible region of wind power is the region below the plane in Fig. \ref{fig:WakeSpeedXi} and above zero, which can be approximated by its convex hull. Multi-Parametric Toolbox (MPT) 3.0 \cite{MPT3} can find the convex hull of a set of points, which is represented by its extreme point set or a linear inequality set describing the facets. We use MPT to find the convex hull of the feasible region of wind power and get a polyhedron as follows.


\begin{align}
\label{eq:region}
    \mathcal{P}^W := \left\{ (x^W,\xi^W,p^W) \middle| 
    \begin{aligned}
    & p^W \geq 0, x^W \leq \bar{x}^W, \xi^W \leq \bar{\xi}^W \\
    & p^W \leq a_k^{(1)} x^W + a_k^{(2)} \xi^W + a_k^{(3)}, \forall k \in \mathcal{K}_0^a
    \end{aligned}
    \right\},
\end{align}
where $a_k^{(1)}$, $a_k^{(2)}$, and $a_k^{(3)}$ are the parameters of inequalities and $\min_{k \in \mathcal{K}_0^a} \{a_k^{(1)} x^W + a_k^{(2)} \xi^W + a_k^{(3)} \}$ is a piecewise linear approximation of the available wind power function.

There may be a large number of inequalities in \eqref{eq:region}. To improve the computational efficiency, we propose Algorithm \ref{alg:1} to select a representative subset of inequalities while maintaining acceptable precision. Algorithm \ref{alg:1} starts from a cuboid containing the original polyhedron, and adds the remaining most important inequality in each iteration until the approximation error is lower than a certain level. Algorithm \ref{alg:1} always converges because the number of inequalities is finite.
\begin{algorithm}
\label{alg:1}
\normalsize
\caption{Selection of representative inequalities}
\begin{algorithmic}[1]
\renewcommand{\algorithmicrequire}{\textbf{Input:}}
\renewcommand{\algorithmicensure}{\textbf{Output:}}
\REQUIRE Error tolerance $\epsilon>0$, $\bar{x}^W$, $\bar{\xi}^W$, and inequality coefficients $(a_k^{(1)},a_k^{(2)},a_k^{(3)}), k \in \mathcal{K}_0^a$.
\ENSURE A set $\mathcal{K}^a \subset \mathcal{K}_0^a$ and the approximate polyhedron $\mathcal{P}$.
\STATE Initiation: Let $\mathcal{K}^a \leftarrow \varnothing$.
\STATE Let the approximate polyhedron $\mathcal{P}$ be
\begin{align}
    \left\{ (x^W,\xi^W,p^W) \middle| 
    \begin{aligned}
    & 0 \leq x^W \leq \bar{x}^W, 0 \leq \xi^W \leq \bar{\xi}^W \\
    & 0 \leq p^W \leq \bar{p}^W \\
    & p^W \leq a_k^{(1)} x^W + a_k^{(2)} \xi^W + a_k^{(3)}, \forall k \in \mathcal{K}^a
    \end{aligned}
    \right\}. \nonumber
\end{align}
\STATE Find the extreme points $ (x_l^W,\xi_l^W,p_l^W),l \in \mathcal{I}$ of $\mathcal{P}$ via MPT, where $\mathcal{I}$ is the index set.
\STATE Calculate the maximum approximation error
\begin{align}
    \max_{l \in \mathcal{I}}\left\{ \max_{k \in \mathcal{K}_0^a\setminus\mathcal{K}^a} \left\{p_l^W - (a_k^{(1)} x_l^W + a_k^{(2)} \xi_l^W + a_k^{(3)}) \right\} \right\}, \nonumber
\end{align}
where the inside max calculates the approximation error at $(x_l^W,\xi_l^W)$. If the error is no larger than $\epsilon$, terminate; otherwise, suppose $k_0 \in \mathcal{K}_0^a\setminus\mathcal{K}^a$ reaches the above maximum, then update $\mathcal{K}^a \leftarrow \mathcal{K}^a \cup \{ k_0 \}$ and go to Step 2.
\end{algorithmic}
\end{algorithm} 

\section{Distributionally Robust Sizing Model}
\label{sec-3}

In this section, we first propose models to evaluate the operation costs given the sizing strategies (capacities), and then build the ambiguity sets, based on which the distributionally robust sizing model is established.

\subsection{Operation Problems}

First, we consider the operation stage with given wind power generation and ESS capacities and uncertainty realizations. Two models are developed to evaluate the operation cost under 1) \emph{normal conditions}, where the total fuel cost is minimized and load shedding is not allowed; and 2) \emph{extreme conditions}, where the total load shedding is minimized.

\subsubsection{Power System Operation Constraints}

Let $S^T := \{1, 2, \dots, T\}$, $S^B$, and $S^L$ be the set of periods, buses, and transmission lines, respectively. Denote the period length by $\Delta_t$. The power system operation constraints are as follows.
\bsq
\label{eq:model}
\begin{align}
    \label{eq:model-1}
    & f_{i,t}^G = \max_{k \in \mathcal{K}^b} \left\{ b_{i,k}^{(1)} p_{i,t}^G \Delta_t + b_{i,k}^{(2)} \right\}, \forall i \in S^B, t \in S^T, \\
    \label{eq:model-2}
    & \underline{P}_i^G \leq p_{i,t}^G \leq \overline{P}_i^G, \forall i \in S^B, t \in S^T, \\
    \label{eq:model-3}
    & \underline{R}_i^G \Delta_t \leq p_{i,t}^G - p_{i,t-1}^G \leq \overline{R}_i^G \Delta_t, \forall i \in S^B, t \in S^T \setminus \{1\}, \\
    \label{eq:model-4}
    & \underline{R}_i^G \Delta_t \leq p_{i,1}^G - p_{i,T}^G \leq \overline{R}_i^G \Delta_t, \forall i \in S^B, \\
    \label{eq:model-5}
    & 0 \leq p_{i,t}^{SC} \leq x_i^{SP}, 0 \leq p_{i,t}^{SD} \leq x_i^{SP}, \forall i \in S^B, t \in S^T, \\
    \label{eq:model-6}
    & e_{i,t}^S = e_{i,t-1}^S + \left(\eta^{SC} p_{i,t}^{SC} - \frac{p_{i,t}^{SD}}{\eta^{SD}} \right) \Delta_t, \forall i \in S^B, t \in S^T \setminus \{1\}, \\
    \label{eq:model-7}
    & e_{i,1}^S = e_{i,T}^S + \left(\eta^{SC} p_{i,1}^{SC} - \frac{p_{i,1}^{SD}}{\eta^{SD}} \right) \Delta_t, \forall i \in S^B, \\
    \label{eq:model-8}
    & \underline{\omega}^S x_i^{SE} \leq e_{i,t}^S \leq \overline{\omega}^S x_i^{SE}, \forall i \in S^B, t \in S^T, \\
    \label{eq:model-9}
    & p_{ij,t}^L = \frac{\theta_{i,t}^B - \theta_{j,t}^B}{X_{ij}^L}, - S_{ij}^L \leq p_{ij,t}^L \leq S_{ij}^L, \forall (i,j) \in S^L, t \in S^T, \\
    \label{eq:model-10}
    & 0 \leq p_{i,t}^D \leq \zeta_{i,t}^D, p_{i,t}^C \geq 0, \forall i \in S^B, t \in S^T, \\
    \label{eq:model-11}
    & \begin{aligned}
    & \sum_{(k,i) \in S^L} p_{ki,t}^L + p_{i,t}^G + \zeta_{i,t}^W + p_{i,t}^{SD} - p_{i,t}^C \\
    = & \sum_{(i,j) \in S^L} p_{ij,t}^L + p_{i,t}^{SC} + \zeta_{i,t}^D - p_{i,t}^D, \forall i \in S^B, t \in S^T.
    \end{aligned}
\end{align}
\esq
Constraints \eqref{eq:model-1}-\eqref{eq:model-4} constitute the model of traditional generators, where $p_{i,t}^G$ and $f_{i,t}^G$ denote the generation power and fuel cost of the traditional generator at bus $i$ in period $t$, respectively. The fuel cost is modeled by the maximum of a set of linear functions in \eqref{eq:model-1}, where $b_{i,k}^{(1)}, b_{i,k}^{(2)}, k \in \mathcal{K}^b$ are constant coefficients. Constraint \eqref{eq:model-2} stipulates the range $[\underline{P}_i^G, \overline{P}_i^G]$ of generation power. Constraints \eqref{eq:model-3} and \eqref{eq:model-4} consist of the lower bound $\underline{R}_i^G$ and upper bound $\overline{R}_i^G$ of ramp rates, where \eqref{eq:model-4} ensures continuous operation. Constraints \eqref{eq:model-5}-\eqref{eq:model-8} compose the model of ESSs, where $p_{i,t}^{SC}$, $p_{i,t}^{SD}$ are the charging and discharging power of the ESS at bus $i$ in period $t$, respectively; $e_{i,t}^S$ is the stored energy at the end of period $t$. The power and energy capacities of the ESS at bus $i$ are denoted by $x_i^{SP}$ and $x_i^{SE}$, respectively. Constraint \eqref{eq:model-5} limits the charging/discharging power. Constraints \eqref{eq:model-6} and \eqref{eq:model-7} are for the dynamics of the stored energy, with charging and discharging efficiencies $\eta^{SC}$ and $\eta^{SD}$. Constraint \eqref{eq:model-8} stipulates the lower bound $\underline{\omega}^S$ and upper bound $\overline{\omega}^S$ of ESSs' state-of-charge. The complementary constraint $p_{i,t}^{SC} p_{i,t}^{SD} = 0$ preventing simultaneous charging and discharging is omitted according to Proposition 1 in \cite{xie2023sizing}, in the sense that the minimized fuel cost or load shedding values will not be affected. The DC power flow model is used in \eqref{eq:model-9}, where $p_{ij,t}^L$ is the active power flow from node $i$ to node $j$ in period $t$, $X_{ij}^L$ and $S_{ij}^L$ are the reactance and capacity of the line, respectively, and $\theta_{i,t}^B$ denotes the voltage angle. The load demand, load shedding, and the power curtailment at bus $i$ in period $t$ are denoted by $\zeta_{i,t}^D$, $p_{i,t}^D$, and $p_{i,t}^C$, respectively. Their bounds are in \eqref{eq:model-10}. Constraint \eqref{eq:model-11} is the nodal power balance equation, in which $\zeta_{i,t}^W$ is the wind power generation at bus $i$ in period $t$.

\subsubsection{Operation Problem in Normal Conditions}
In normal conditions, the goal is to minimize the total fuel cost of traditional generators.
\bsq
\label{eq:operation-fuel}
\begin{align}
    \label{eq:operation-fuel-1}
    & g^N(x, \zeta) := \min_{f^G, p^G, p^{SC}, p^{SD}, e^S, p^L, \theta^B, p^C} \sum_{t \in S^T} \sum_{i \in S^B} f_{i,t}^G \\
    \label{eq:operation-fuel-2}
    \mbox{s.t.}~ & f_{i,t}^G \geq b_{i,k}^{(1)} p_{i,t}^G \Delta_t + b_{i,k}^{(2)}, \forall k \in \mathcal{K}^b, i \in S^B, t \in S^T \\
    \label{eq:operation-fuel-3}
    & p_{i,t}^D = 0, \forall i \in S^B, t \in S^T, \eqref{eq:model-2}-\eqref{eq:model-11}.
\end{align}
\esq
The objective \eqref{eq:operation-fuel-1} minimizes the total fuel cost. The fuel cost model in \eqref{eq:model-1} can be equivalently transformed into inequalities in \eqref{eq:operation-fuel-2} thanks to the minimization operation in the objective. The remaining constraints in the power system model are included and the load shedding is forced to be zero in \eqref{eq:operation-fuel-3} since no load shedding is allowed in normal conditions. Let $x := (x_i^W, x_i^{SP}, x_i^{SE}; i \in S^B)$ denote the capacity variable and $\zeta := (\zeta_{i,t}^W, \zeta_{i,t}^D; i \in S^B, t \in S^T)$ be the uncertainty realization. When $(x,\zeta)$ changes, the optimal objective value \eqref{eq:operation-fuel-1} changes accordingly, which is denoted by $g^N(x,\zeta)$.

\subsubsection{Operation Problem in Extreme Conditions}
In extreme conditions, the goal is to minimize the total load shedding.
\bsq
\label{eq:operation-load}
\begin{align}
    \label{eq:operation-load-1}
    g^E(x, \zeta) := \min_{p^D, p^G, p^{SC}, p^{SD}, e^S, p^L, \theta^B, p^C}~ & \sum_{t \in S^T} \sum_{i \in S^B} p_{i,t}^D \Delta_t \\
    \label{eq:operation-load-2}
    \mbox{s.t.}~ & \eqref{eq:model-2}-\eqref{eq:model-11}.
\end{align}
\esq
The total load shedding in \eqref{eq:operation-load-1} is minimized subject to the power system constraints in \eqref{eq:operation-load-2}.  Fuel cost is not considered since it is not a major concern in extreme conditions. 
Similarly, the optimal objective value is a function of $x$ and $\zeta$ denoted by $g^E(x, \zeta)$, showing the total load shedding under given $(x,\zeta)$.

\subsection{Decision-Dependent Ambiguity Sets}
We use ambiguity sets to model uncertainties, formed by a set of distributions around the empirical distribution with distances measured by Wasserstein metric.

\subsubsection{Empirical Distributions}

Suppose we have some empirical data $(\hat{\xi}_{i,t,n}^W, \hat{\zeta}_{i,t,n}^D; i \in S^B, t \in S^T), n \in S^E$ of wind and load demand in extreme conditions, where $S^E$ is the index set of extreme scenarios. Then the available wind power is
\begin{align}
\label{eq:available-wind}
    \hat{\zeta}_{i,t,n}^W (x_i^W) = \min \left\{  \min_{k \in \mathcal{K}^a} \{ a_{i,k}^{(1)} x_i^W + a_{i,k}^{(2)} \hat{\xi}_{i,t,n}^W + a_{i,k}^{(3)} \}, \hat{\xi}_{i,t,n}^W x_i^W \right\},
\end{align}
where $\min_{k \in \mathcal{K}^a} \{ a_{i,k}^{(1)} x_i^W + a_{i,k}^{(2)} \hat{\xi}_{i,t,n}^W + a_{i,k}^{(3)} \}$ is the linearized approximation obtained by Algorithm \ref{alg:1}; $\hat{\xi}_{i,t,n}^W x_i^W$ is a theoretical upper bound for the available wind power without considering the wake effect and the cut-out wind speed, which can be seen in \eqref{eq:pw} and \eqref{eq:xi}.
The notation $\hat{\zeta}_{i,t,n}^W (x_i^W)$ is to emphasize that the available wind power is decision-dependent. Thus, the data of scenario $n \in S^E$ are $\hat{\zeta}_n(x) := (\hat{\zeta}_{i,t,n}^W (x_i^W),\hat{\zeta}_{i,t,n}^D; i \in S^B, t \in S^T)$. Based on the data, the empirical probability distribution $\mathbb{P}_0^E (x)$ in extreme conditions is
\begin{align}
    \mathbb{P}_0^E (x) := \frac{1}{|S^E|} \sum_{n \in S^E} \delta_{\hat{\zeta}_n (x)} = \frac{1}{|S^E|} \sum_{n \in S^E} \left( \bigotimes_{i \in S^B} \delta_{\hat{\zeta}_{i,n}^W (x_i^W)} \right) \otimes \delta_{\hat{\zeta}_n^D}, \nonumber
\end{align}
where $|S^E|$ is the number of scenarios in $S^E$, $\delta$ denotes the Dirac distribution with unit mass at the subscript point, and $\otimes$ represents the product operation of distributions. Therefore, probability $1/|S^E|$ is assigned to each available wind power sample $\hat{\zeta}_n(x)$ under the empirical distribution $\mathbb{P}_0^E(x)$. Denote the index set of normal scenarios by $S^N$, then similarly the corresponding empirical distribution $\mathbb{P}_0^N(x)$ can be defined.

\subsubsection{Range of Uncertain Variables}

For the wind farm at bus $i$, the range $\Xi_i^{WE} (x_i^W)$ of the wind power output $\zeta_i^W$ in extreme conditions is built by
\begin{align}
    \Xi_i^{WE} (x_i^W) := \left\{ \zeta_i^W \middle| 
    \begin{aligned}
    & \zeta_{i,t}^W \leq a_{i,k}^{(1)} x_i^W + a_{i,k}^{(2)} \overline{\xi}_i^W + a_{i,k}^{(3)}, \forall k \in \mathcal{K}^a \\
    & 0 \leq \zeta_{i,t}^W \leq x_i^W
    \end{aligned}
     \right\}. \nonumber
\end{align}
The wind power $\zeta_{i,t}^W$ cannot exceed the bounds obtained in Algorithm 1 as well as the capacity $x_i^W$. Again we denote the wind power range by $\Xi_i^{WE} (x_i^W)$ to emphasize the decision-dependent feature.

The range of load demand in extreme conditions is given by
$
    \Xi^{DE} := \{ \zeta^D | 0 \leq \zeta_{i,t}^D \leq \overline{\zeta}_i^D \}
$,
where $\overline{\zeta}_i^D$ is the maximum possible load demand at bus $i$.

For normal conditions, the range is defined as the convex hull of the corresponding scenarios. Then the range of wind power $\zeta_i^W$ in normal conditions is
\begin{align}
    \Xi_i^{WN} (x_i^W) := \left\{ \zeta_i^W = \sum_{n \in S^N} \lambda_n \hat{\zeta}_{i,n}^W (x_i^W) \middle| \lambda_n \geq 0, \sum_{n \in S^N} \lambda_n = 1 \right\}, \nonumber
\end{align}
where $\lambda_n, n \in S^N$ are the coefficients of convex combination; all the convex combinations of $\hat{\zeta}_{i,n}^W (x_i^W), n \in S^N$ are contained in $\Xi_i^{WN}(x_i^W)$. The range $\Xi^{DN}$ of load demand in normal conditions has a similar form. Compared with normal conditions, the range of uncertain variables in extreme conditions is larger to include the possible extreme scenarios, i.e., $\Xi_i^{WN} (x_i^W) \subset \Xi_i^{WE} (x_i^W)$ and $\Xi^{DN} \subset \Xi^{DE}$.

\subsubsection{Wasserstein Metric and Ambiguity Set}

Wasserstein metric is a way to measure the distance between two distributions \cite{kantorovich1958space}. For distributions $\mathbb{P}_1$ and $\mathbb{P}_2$ defined on $\Xi$, the 1-norm Wasserstein metric between them is
\begin{align}
    & d^W(\mathbb{P}_1, \mathbb{P}_2) := \inf_\Pi \int_{\Xi^2} \| \zeta_1 - \zeta_2 \|_1 \Pi(d\zeta_1, d\zeta_2) \\
    & \Pi ~\mbox{is a joint distribution with marginals}~ \mathbb{P}_1 ~\mbox{and}~ \mathbb{P}_2, \nonumber
\end{align}
where $\|\cdot\|_1$ is the 1-norm. It can be regarded as the minimum cost of transporting the mass distribution $\mathbb{P}_1$ to $\mathbb{P}_2$, in which the transport way is represented by the joint distribution $\Pi$. 

Inspired by the definition of Wasserstein metric, we propose the following ambiguity set $\mathcal{B}^E (\epsilon^E, x)$ for uncertain variable $\zeta$ in extreme conditions.
\begin{align}
\label{eq:ambiguity}
    & \mathcal{B}^E (\epsilon^E, x) := \left\{
    \frac{1}{|S^E|} \sum_{n \in S^E} \left( \bigotimes_{i \in S^B}  \mathbb{P}_{i,n}^W \right) \otimes \mathbb{P}_n^D \middle| \right. \nonumber \\
    & \mathbb{P}_{i,n}^W \in \mathcal{M} (\Xi_i^{WE} (x_i^W)), \Pi_i^W = \frac{1}{|S^E|} \sum_{n \in S^E} \mathbb{P}_{i,n}^W \otimes \delta_{\hat{\zeta}_{i,n}^W (x_i^W)}, \nonumber \\
    & \int_{(\Xi_i^{WE} (x_i^W))^2} \|\zeta_i^W - \tilde{\zeta}_i^W \|_1 \Pi_i^W(d\zeta_i^W, d\tilde{\zeta}_i^W) \leq \epsilon_i^{WE} x_i^W, \forall i \in S^B, \nonumber \\
    & \mathbb{P}_n^D \in \mathcal{M}(\Xi^{DE}), \Pi^D = \frac{1}{|S^E|} \sum_{n \in S^E} \mathbb{P}_n^D \otimes \delta_{\hat{\zeta}_n^D}, \nonumber \\
    & \left. \int_{(\Xi^{DE})^2} \| \zeta^D - \tilde{\zeta}^D \|_1 \Pi^D (d\zeta^D, d\tilde{\zeta}^D) \leq \epsilon^{DE} \right\}.
\end{align}
The distributions in $\mathcal{B}^E (\epsilon^E, x)$ are the average of the product distributions of $\mathbb{P}_{i,n}^W$ and $\mathbb{P}_n^D$. $\mathcal{M}(\cdot)$ denotes the set of all the distributions supported on the range in the bracket. $\Pi_i^W$ is a joint distribution with marginals $\frac{1}{|S^E|} \sum_{n \in S^E} \mathbb{P}_{i,n}^W$ and the empirical distribution $\frac{1}{|S^E|} \sum_{n \in S^E} \delta_{\hat{\zeta}_{i,n}^W (x_i^W)}$. The Wasserstein metric between them is no larger than $\epsilon_i^{WE} x_i^W$, where $\epsilon_i^{WE}$ is a parameter controlling the size of the ambiguity set. Since larger capacity leads to larger fluctuation of wind power, we use the product $\epsilon_i^{WE} x_i^W$ as the bound. The constraints of load demand distribution $\mathbb{P}_n^D$ have a similar bound $\epsilon^{DE}$. Let $\epsilon^E := (\epsilon_i^{WE}, \epsilon^{DE}; i \in S^E)$, then the ambiguity set contains $\mathbb{P}_0^E(x)$ for all non-negative $\epsilon^E$ and it becomes a singleton when $\epsilon^E = 0$, the zero vector.

The ambiguity set $\mathcal{B}^N (\epsilon^N, x)$ in normal conditions can be defined similarly by replacing $S^E$, $\Xi_i^{WE}$, $\Xi^{DE}$, $\epsilon_i^{WE}$, and $\epsilon^{DE}$ by $S^N$, $\Xi_i^{WN}$, $\Xi^{DN}$, $\epsilon_i^{WN}$, and $\epsilon^{DN}$, respectively. Both the ambiguity sets are decision-dependent.

There are theoretical results about the convergence rate of empirical distributions under Wasserstein metric \cite{fournier2015on}, which guide the proper selection of the Wasserstein metric bounds $\epsilon^E$ and $\epsilon^N$. Roughly speaking, the bound should be proportional to $\sup_{\zeta_1,\zeta_2 \in \Xi} \| \zeta_1 - \zeta_2 \|_1$ and $1/N^{1/M}$, where $N$ is the number of data and $M$ is the dimension.

Some literature on Wasserstein-metric-based DRO uses the ambiguity sets that only contain the discrete distributions supported on the sample data sets \cite{luo2020distributionally}, which are not suitable for wind power and load demand uncertainties because they can change continuously. In contrast, the Wasserstein-metric-based ambiguity sets constructed in this paper contain general distributions including both continuous and discrete ones.

\subsection{Sizing Model}

Suppose $\overline{x}_i^W$, $\overline{x}_i^{SP}$, and $\overline{x}_i^{SE}$ are the upper bounds of $x_i^W$, $x_i^{SP}$, and $x_i^{SE}$, respectively. Then the feasible set of capacity variable $x$ is as follows.
\begin{align}
    \mathcal{X} := \left\{ x \middle| 
    \begin{aligned}
    & 0 \leq x_i^W \leq \overline{x}_i^W, 0 \leq x_i^{SP} \leq \overline{x}_i^{SP}, 0 \leq x_i^{SE} \leq \overline{x}_i^{SE}, \forall i \in S^B \\
    & g^N(x, \hat{\zeta}_n) < +\infty, \forall n \in S^N
    \end{aligned}
    \right\}, \nonumber
\end{align}
where $g^N(x, \hat{\zeta}_n) < +\infty, \forall n \in S^N$ means that the operation problem \eqref{eq:operation-fuel} is feasible for all normal scenarios; in other words, load shedding will not arise in normal conditions. Since $g^N(x, \zeta)$ is the optimal value of a linear program (LP), it is a convex function on its domain \cite{boyd2004convex}.

The bi-objective sizing model is established below.
\bsq
\label{eq:planning}
\begin{align}
    \label{eq:planning-1}
    \min_{x \in \mathcal{X}}~ & \left\{\sum_{i \in S^B} (C_i^W x_i^W + C^{SP} x_i^{SP} + C^{SE} x_i^{SE}), I^F \right\} \\
    \label{eq:planning-2}
    \mbox{s.t.}~ & \sup_{\mathbb{P} \in \mathcal{B}^E (\epsilon^E, x)} \mathbb{E}_\mathbb{P} [g^E(x,\zeta)] \leq \overline{g}^E \\
    \label{eq:planning-3}
    & I^F = \sup_{\mathbb{P} \in \mathcal{B}^N (\epsilon^N, x)} \mathbb{E}_\mathbb{P} [g^N(x,\zeta)],
\end{align}
\esq
where in \eqref{eq:planning-1} the first objective is to minimize the total investment cost, and the second objective function defined in \eqref{eq:planning-3} represents the worst-case fuel cost expectation when the distribution $\mathbb{P}$ varies within the ambiguity set $\mathcal{B}^N (\epsilon^N, x)$ for normal conditions. Constraint \eqref{eq:planning-2} stipulates an upper bound $\overline{g}^E$ for the worst-case load shedding expectation in extreme conditions. The worst-case distributions in normal and extreme conditions are different. In \eqref{eq:planning}, the wind farm capacities are approximated by continuous variables. If the obtained capacity is not divisible by the capacity of a single WT, we round the obtained capacity to the nearest integer multiples of the WT capacity to ensure the number of WTs is an integer.
For large-scale wind farms, this simplification would not significantly affect the optimality, as will be examined in Section \ref{sec-5}.

The sizing model \eqref{eq:planning} is hard to solve due to three reasons: 1) There are distributionally robust expectations in the model. 2) The ambiguity sets are decision-dependent. 3) The functions $g^E(x,\zeta)$ and $g^N(x,\zeta)$ are defined as the optimal values of LP problems. To overcome these difficulties, we develop a solution method in the next section.

\section{Solution Method}
\label{sec-4}

We first transform the DRO sizing model \eqref{eq:planning} into a SP problem, and then propose an iterative algorithm to solve it.

\subsection{Approximate Stochastic Programming Problem}

First, we use Lipschitz constants to transform the distributionally robust expectations into the expectations with respect to the empirical distributions plus compensation terms, as in the following theorem.

\begin{theorem}
\label{thm:1}
For any $x \in \mathcal{X}$ and $\epsilon^E \geq 0$,
\begin{align}
\label{eq:DR-expectation}
    & \sup_{\mathbb{P} \in \mathcal{B}^E (\epsilon^E, x)} \mathbb{E}_\mathbb{P} [g^E(x,\zeta)] \nonumber \\
    \leq~ & \sum_{i \in S^B} \epsilon_i^{WE} x_i^W L_i^{WE}(x) + \epsilon^{DE}L^{DE}(x) + \mathbb{E}_{\mathbb{P}_0^E (x)} [g^E(x,\zeta)],
\end{align}
where $L_i^{WE}(x)$ is the minimum Lipschitz constant of $g^E(x,\zeta)$ regarded as a function of $\zeta_i^W$, and $L^{DE}(x)$ is the minimum Lipschitz constant of $g^E(x,\zeta)$ regarded as a function of $\zeta^D$. \footnote{For example,
\begin{align}
    L_i^{WE}(x) := \inf \left\{ L ~\middle|~ 
    \begin{aligned}
    & \left| g^E(x,\zeta) - g^E(x, \tilde{\zeta}) \right| \leq L \left\| \zeta_i^W - \tilde{\zeta}_i^W \right\|_1 \\
    & \forall \zeta_i^W, \tilde{\zeta}_i^W \in \Xi_i^{WE}(x_i^W), \mbox{s.t.}~ \zeta^D = \tilde{\zeta}^D \in \Xi^{DE} \\
    & \zeta_j^W = \tilde{\zeta}_j^W \in \Xi_j^{WE}(x_j^W), \forall j \neq i \\
    \end{aligned}
    \right\}. \nonumber
\end{align}}
\end{theorem}

The proof of Theorem \ref{thm:1} can be found in Appendix \ref{appendix-A}. The distributionally robust expectation in normal conditions has a similar property with minimum Lipschitz constants $L_i^{WN}(x)$ and $L^{DN}(x)$. 

By the sample average approximation (SAA) technique, $\mathbb{E}_{\mathbb{P}_0^E (x)} [g^E(x,\zeta)] = \sum_{n \in S^E} g^E(x, \hat{\zeta}_n(x)) / |S^E|$. Thus, the following SP problem is an approximation of the sizing model \eqref{eq:planning}. 
\bsq
\label{eq:SP}
\begin{align}
    \label{eq:SP-1}
    \min~ & \left\{\sum_{i \in S^B} (C_i^W x_i^W + C^{SP} x_i^{SP} + C^{SE} x_i^{SE}), I^F \right\} \\
    \label{eq:SP-2}
    \mbox{s.t.}~ & 0 \leq x_i^W \leq \overline{x}_i^W, 0 \leq x_i^{SP} \leq \overline{x}_i^{SP}, 0 \leq x_i^{SE} \leq \overline{x}_i^{SE}, \forall i \in S^B \\
    \label{eq:SP-3}
    & \sum_{i \in S^B} \epsilon_i^{WE} x_i^W L_i^{WE}(x) + \epsilon^{DE} L^{DE}(x) + \frac{1}{|S^E|} \sum_{n \in S^E} g_n^E \leq \overline{g}^E \\
    \label{eq:SP-4}
    & g_n^E \geq \sum_{t \in S^T} \sum_{i \in S^B} p_{i,t,n}^D \Delta_t, \eqref{eq:operation-load-2}, \forall n \in S^E \\
    \label{eq:SP-5}
    & I^F = \sum_{i \in S^B} \epsilon_i^{WN} x_i^W L_i^{WN}(x) + \epsilon^{DN} L^{DN}(x) + \frac{1}{|S^N|} \sum_{n \in S^N} g_n^N \\
    \label{eq:SP-6}
    & g_n^N \geq \sum_{t \in S^T} \sum_{i \in S^B} f_{i,t,n}^G, \eqref{eq:operation-fuel-2}, \eqref{eq:operation-fuel-3}, \forall n \in S^N \\
    \label{eq:SP-7}
    & \zeta_{i,t,n}^W \leq a_{i,k}^{(1)} x_i^W + a_{i,k}^{(2)} \hat{\xi}_{i,t,n}^W + a_{i,k}^{(3)}, \forall k \in \mathcal{K}^a, n \in S^E \cup S^N \\
    \label{eq:SP-8}
    & 0 \leq \zeta_{i,t,n}^W \leq \hat{\xi}_{i,t,n}^W x_i^W, \forall i \in S^B, n \in S^E \cup S^N,
\end{align}
\esq
where \eqref{eq:SP-2} and \eqref{eq:SP-6} contain the requirement $x \in \mathcal{X}$; $g_n^E$ and $g_n^N$ are auxiliary variables representing $g^E(x,\zeta_n(x))$ and $g^N(x,\zeta_n(x))$, respectively. As will be examined in Section \ref{sec-5}, the approximate problem is not over-conservative with suitable parameters $\epsilon^E$ and $\epsilon^N$.

\subsection{Solution Algorithm}

To solve the SP problem \eqref{eq:SP}, we still need to address two issues: 1) what are the minimum Lipschitz constants $L_i^{WE}(x)$, $L^{DE}(x)$, $L_i^{WN}(x)$, and $L^{DN}(x)$; 2) how to solve the bi-objective optimization. 

For the first issue, though the minimum Lipschitz constants are hard to calculate accurately, it is easy for us to obtain their upper bounds based on the fact that functions $g^E(x, \zeta)$ and $g^N(x, \zeta)$ are defined by LPs whose dual variables can reflect their changing rates in $\zeta$. Denote the upper bounds by $\tilde{L}_i^{WE}(x)$, $\tilde{L}^{DE}(x)$, $\tilde{L}_i^{WN}(x)$, and $\tilde{L}^{DN}(x)$, respectively. The detailed procedure for getting the upper bounds is presented in Appendix \ref{appendix-B}. According to the procedure, we can find that calculating $\tilde{L}_i^{WE}(x)$ and $\tilde{L}^{DE}(x)$ involves solving a KKT condition embedded optimization while calculating $\tilde{L}_i^{WN}(x)$ and $\tilde{L}^{DN}(x)$ only requires solving some LP problems. To accelerate the computation, we use uniform upper bounds for ${L}_i^{WE}(x)$ and ${L}^{DE}(x)$, i.e., $\tilde{L}_i^{WE} \geq L_i^{WE}(x), \tilde{L}^{DE} \geq L^{DE}(x), \forall x \in \mathcal{X}$; and $x$-dependent upper bounds for $\tilde{L}_i^{WN}(x)$ and $\tilde{L}^{DN}(x)$, i.e., $\tilde{L}_i^{WN}(x) \geq L_i^{WN}(x)$ and $\tilde{L}^{DN}(x) \geq L^{DN}(x)$.


For the second issue, we adopt the $\epsilon$-constraint method \cite{mavalizadeh2021robust} to handle the two objectives. Particularly, the investment cost objective is replaced with a budget constraint, and the new model only minimizes the fuel cost expectation as follows. 
\begin{align}
    \label{eq:budget}
    \min~ & I^F \nonumber \\
    \mbox{s.t.}~ & \eqref{eq:SP-2}-\eqref{eq:SP-8}\nonumber \\
    & \sum_{i \in S^B} (C_i^W x_i^W + C^{SP} x_i^{SP} + C^{SE} x_i^{SE}) \leq \overline{C},
\end{align}
where $\overline{C}$ is the budget. With a varying budget, the Pareto frontier of the bi-objective sizing model can be obtained, which is the advantage of the $\epsilon$-constraint method compared with summing the two objective functions together. The uniform upper bounds $\tilde{L}_i^{WE}$ and $\tilde{L}^{DE}$ are calculated in advance and used in the solution process of \eqref{eq:budget}, which is shown in Algorithm \ref{alg:2}. The algorithm finds an approximate solution in an iterative way: in each iteration $k$, use the $\tilde{L}_i^{WN}(x_k)$ and $\tilde{L}^{DN}(x_k)$ in the current iteration to approximate the upper bounds of  minimum Lipschitz constants in the next iteration by letting $L_i^{WN} \leftarrow \tilde{L}_i^{WN}(x_k)$, $\forall i \in S^B$, $L^{DN} \leftarrow \tilde{L}^{DN}(x_k)$.
\begin{algorithm}
\label{alg:2}
\normalsize
\caption{Iterative Algorithm for Problem \eqref{eq:budget}}
\begin{algorithmic}[1]
\renewcommand{\algorithmicrequire}{\textbf{Input:}}
\renewcommand{\algorithmicensure}{\textbf{Output:}}
\REQUIRE Parameters of problem \eqref{eq:planning}, budget $\overline{C}$, error tolerance $\epsilon > 0$, and uniform upper bounds $\tilde L_i^{WE}$ and $\tilde L^{DE}$.
\ENSURE Solution $x_k$ and the corresponding distributionally robust expectation of fuel cost $I^F$.
\STATE Initiation: Let $L_i^{WN} \leftarrow 0, \forall i \in S^B$, $L^{DN} \leftarrow 0$, $k \leftarrow 0$, and $x_k \leftarrow 0$. 
\STATE Solve the LP problem
\begin{align}
    \min~ & \sum_{i \in S^B} \epsilon_i^{WN} x_i^W L_i^{WN} + \epsilon^{DN} L^{DN} + \frac{1}{|S^N|} \sum_{n \in S^N} g_n^N \nonumber \\
    \mbox{s.t.}~ & \sum_{i \in S^B} (C_i^W x_i^W + C^{SP} x_i^{SP} + C^{SE} x_i^{SE}) \leq \overline{C} \nonumber \\
    & \sum_{i \in S^B} \epsilon_i^{WE} x_i^W \tilde{L}_i^{WE} + \epsilon^{DE} \tilde{L}^{DE} + \frac{1}{|S^E|} \sum_{n \in S^E} g_n^E \leq \overline{g}^E \nonumber \\
    & \eqref{eq:SP-2}, \eqref{eq:SP-4}, \eqref{eq:SP-6}-\eqref{eq:SP-8}. \nonumber
\end{align}
If it is infeasible, terminate. Otherwise, update $k \leftarrow k + 1$; let $x_k$ be the optimal solution and $I^F$ be the optimal value.
\STATE If $k > 0$ and $\|x_k - x_{k-1}\| \leq \epsilon$, output results and terminate. Otherwise, calculate and let $L_i^{WN} \leftarrow \tilde{L}_i^{WN}(x_k)$, $\forall i \in S^B$, $L^{DN} \leftarrow \tilde{L}^{DN}(x_k)$, then go to Step 2.
\end{algorithmic}
\end{algorithm} 

\section{Case Studies}
\label{sec-5}

This section first uses a modified IEEE 30-bus system to test the proposed method. The influence of the wake effect on the sizing results is examined and the proposed method is compared with the traditional SP and RO methods. Sensitivity analysis is also conducted. Moreover, the scalability of the proposed method is verified via a modified IEEE 118-bus system. All experiments are performed on a laptop equipped with Intel i7-12700H processor and 16 GM RAM. The LPs and mixed-integer linear programs (MILPs) are solved by GUROBI 9.5 in MATLAB.

\subsection{Benchmark}

The benchmark case study is based on a modified IEEE 30-bus system, whose data can be found in \cite{xie2022github} and some parameters are listed in TABLE \ref{table:parameter}. There are candidate wind farms at buses 13 and 27, where the WT parameters are from \cite{yang2019wind}. The candidate wind farms are built on square areas with 25 and 20 rows, respectively. The wind speed data are generated based on the weather data in Qinghai province, China \cite{staffell2016using}. There are candidate ESSs at buses 13, 23, and 27. They all have capacity bounds $\overline{x}^{SP} = 1$ GW and $\overline{x}^{SE} = 5$ GWh. The load demand data are generated by normal distributions with bounds.
We generate 365 groups of data in total. By analyzing the load demand and wind conditions, we select 89 groups from them to represent extreme conditions, and another 272 groups are for normal conditions. In the benchmark case, $|S^E| = 24$ and $|S^N| = 72$ groups of data are used for sizing. In the out-of-sample tests, $65$ and $200$ groups of data are used for validation, representing the distributions of extreme conditions and normal conditions, respectively. The parameters $\epsilon^E$ and $\epsilon^N$ controlling the size of the ambiguity sets are set as
\begin{align}
    \epsilon_i^{WE} = \epsilon_0 / |S^E|^{1/T}, i = 13, 27, \epsilon^{DE} = 20 \epsilon_0 / |S^E|^{1/(20T)}, \nonumber \\
    \epsilon_i^{WN} = \epsilon_0 / |S^N|^{1/T}, i = 13, 27, \epsilon^{DN} = 20 \epsilon_0 / |S^N|^{1/(20T)}, \nonumber
\end{align}
where 20 is the number of buses with load, and $\epsilon_0 = 0.05$ is an auxiliary scalar parameter to control $\epsilon^E$ and $\epsilon^N$.

\begin{table}[!t]
\scriptsize
\renewcommand{\arraystretch}{1.3}
\caption{Parameters}
\label{table:parameter}
\centering
\begin{tabular}{cc|cc}
\hline
Parameter & Value & Parameter & Value \\
\hline
$T$ & 24 & $\Delta_t$ & $1$ h \\
$\eta^{SC}$ & $0.95$ & $\eta^{SD}$ & $0.95$ \\
$\underline{\omega}^S$ & $0.10$ & $\overline{\omega}^S$ & $0.90$ \\
$C^{SP}$ & $1.0 \times 10^6$ CNY/MW & $C^{SE}$ & $1.2 \times 10^6$ CNY/MWh\\
$C^W$ & $5.5 \times 10^6$ CNY/MW & $\overline{g}^E$ & $220$ MWh \\
\hline
\end{tabular}
\end{table}

\begin{figure}[t]
\centering
\includegraphics[width=1.0\columnwidth]{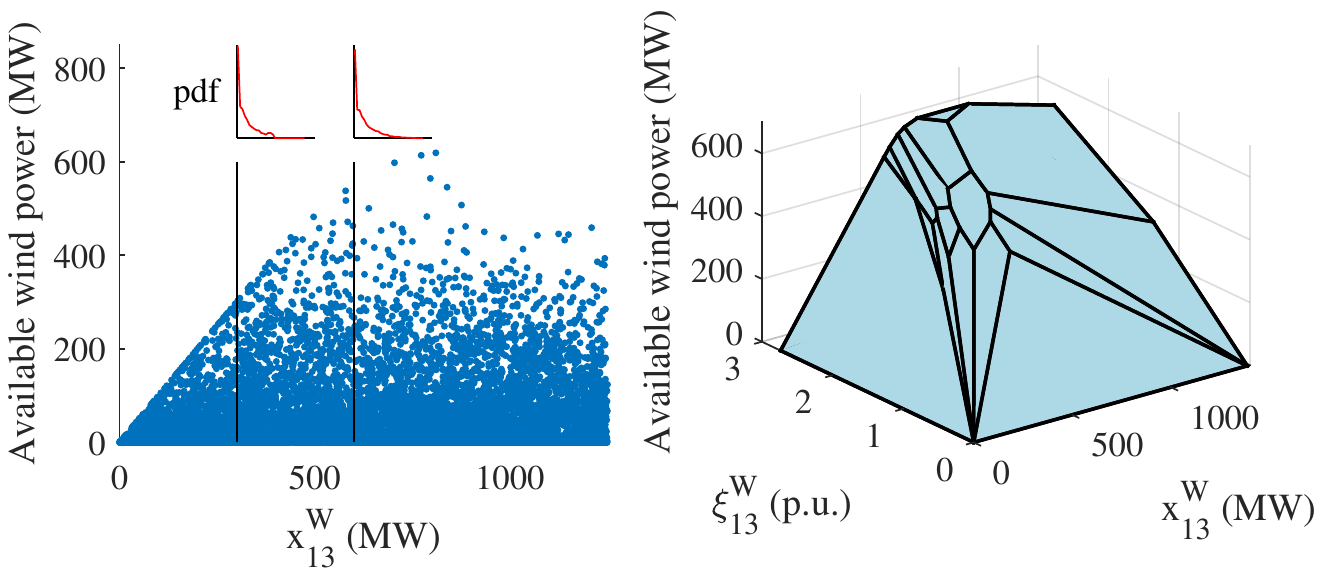}
\caption{Available wind power samples and linearized available wind power function at bus 13.}
\label{fig:Linear}
\end{figure}

To visualize the endogenous uncertainty, we draw the available wind power sample points in Fig. \ref{fig:Linear} (left), where the points with the same horizontal coordinate reflect the distribution and probability density function (pdf) of the available wind power under a specific wind power capacity. It is clear that the distributions change as the capacity varies. The linearized available wind power function proposed in Section \ref{sec-2} is shown in Fig. \ref{fig:Linear} (right). The total computation time is within 1 s. The errors between the available wind power model \eqref{eq:available-wind} and the simulation value are about 5\% on average, mainly from the convex hull approximation. 

\begin{figure}[t]
\centering
\includegraphics[width=0.7\columnwidth]{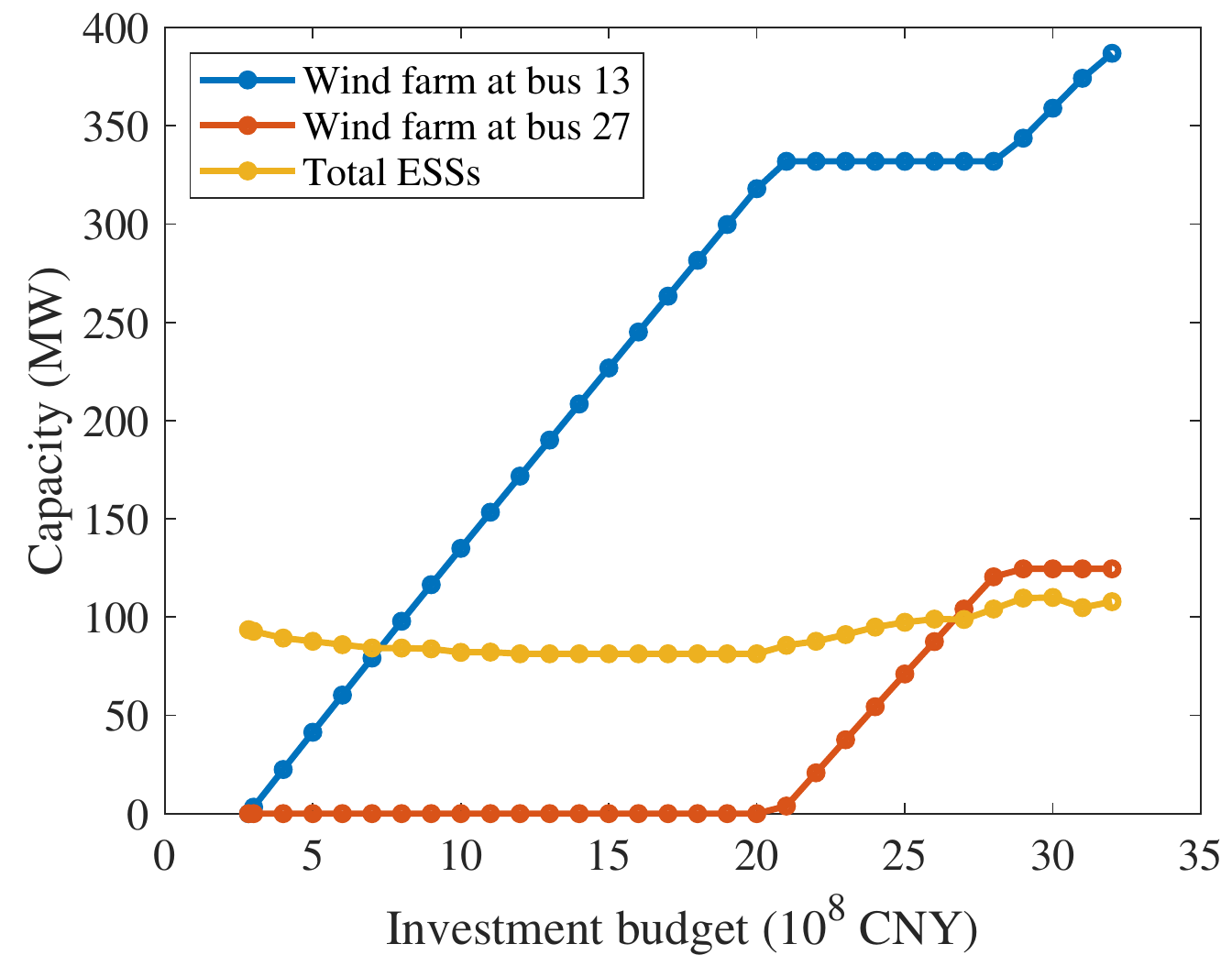}
\caption{Sizing results of the modified IEEE 30-bus system.}
\label{fig:Capacity}
\end{figure}

The sizing results with different investment budgets are shown in Fig. \ref{fig:Capacity}. The minimum investment is $2.83 \times 10^8$ CNY to equip ESSs and ensure the reliable supply of load demand. When the investment budget increases, the capacities of the wind power generation become larger because renewable power can help reduce fuel costs. The wind power generation capacity accounts for about $30$\% of the total generation capacity when the investment budget is $32 \times 10^8$ CNY. The total capacity of ESSs does not change much when wind power penetration is relatively low, but starts to grow with more wind power generation due to the flexibility requirement in the power system. We compare the results before and after the obtained capacity is rounded to integer multiples of the single WT capacity, and the error is below 0.5\%.
Therefore, we can optimize with continuous capacity variables and then round the results without perceptible loss of optimality. Moreover, all fuel costs and load shedding in this section have been verified to remain the same after adding the charging and discharging complementarity constraints to the power system model \eqref{eq:model}.


\subsection{Comparison With Traditional Methods}

In order to show the impact of the wake effect and the advantage of the DRO technique, we compare the proposed method with traditional methods:
\begin{itemize}
    \item DRO: The proposed method.
    \item SP1: Traditional SP method that does not consider the wake effect. In this method, the empirical distributions are regarded as the real distributions, i.e., $\sup_{\mathbb{P} \in \mathcal{B}^E (\epsilon^E, x)} \mathbb{E}_\mathbb{P} [g^E (x, \zeta)]$ and $\sup_{\mathbb{P} \in \mathcal{B}^N (\epsilon^N, x)} \mathbb{E}_\mathbb{P} [g^N (x, \zeta)]$ are replaced by $\mathbb{E}_{\mathbb{P}_0^E(x)} [g^E (x, \zeta)]$ and $\mathbb{E}_{\mathbb{P}_0^N(x)} [g^N (x, \zeta)]$, respectively. The wake effect is not considered, so the available wind power in \eqref{eq:available-wind} becomes $\hat{\zeta}_{i,t,n}^W (x_i^W) = \hat{\xi}_{i,t,n}^W x_i^W$. The sizing model can be equivalently transformed into an LP following a similar solution method to that described in Section \ref{sec-4}.
    \item SP2: Traditional SP method considering the wake effect. This method is similar to SP1 except that the available wind power is still computed according to \eqref{eq:available-wind}.
    \item RO: RO method considering the wake effect. This method optimizes over the worst cases in $\tilde{\Xi}^E (x) := \{ \hat{\zeta}_n (x) | n \in S^E \}$ and $\tilde{\Xi}^N (x) := \{ \hat{\zeta}_n (x) | n \in S^N \}$. The sizing model is
    \begin{align}
    \min_{x \in \mathcal{X}}~ & \left\{ \sum_{i \in S^B} (C_i^W x_i^W + C^{SP} x_i^{SP} + C^{SE} x_i^{SE}), \sup_{n \in S^N} g^N(x, \hat{\zeta}_n (x)) \right\} \nonumber \\
    \mbox{s.t.}~ & \sup_{n \in S^E} g^E(x, \hat{\zeta}_n (x)) \leq \overline{g}^E. \nonumber
    \end{align}
\end{itemize}

\begin{figure}[t]
\centering
\includegraphics[width=0.7\columnwidth]{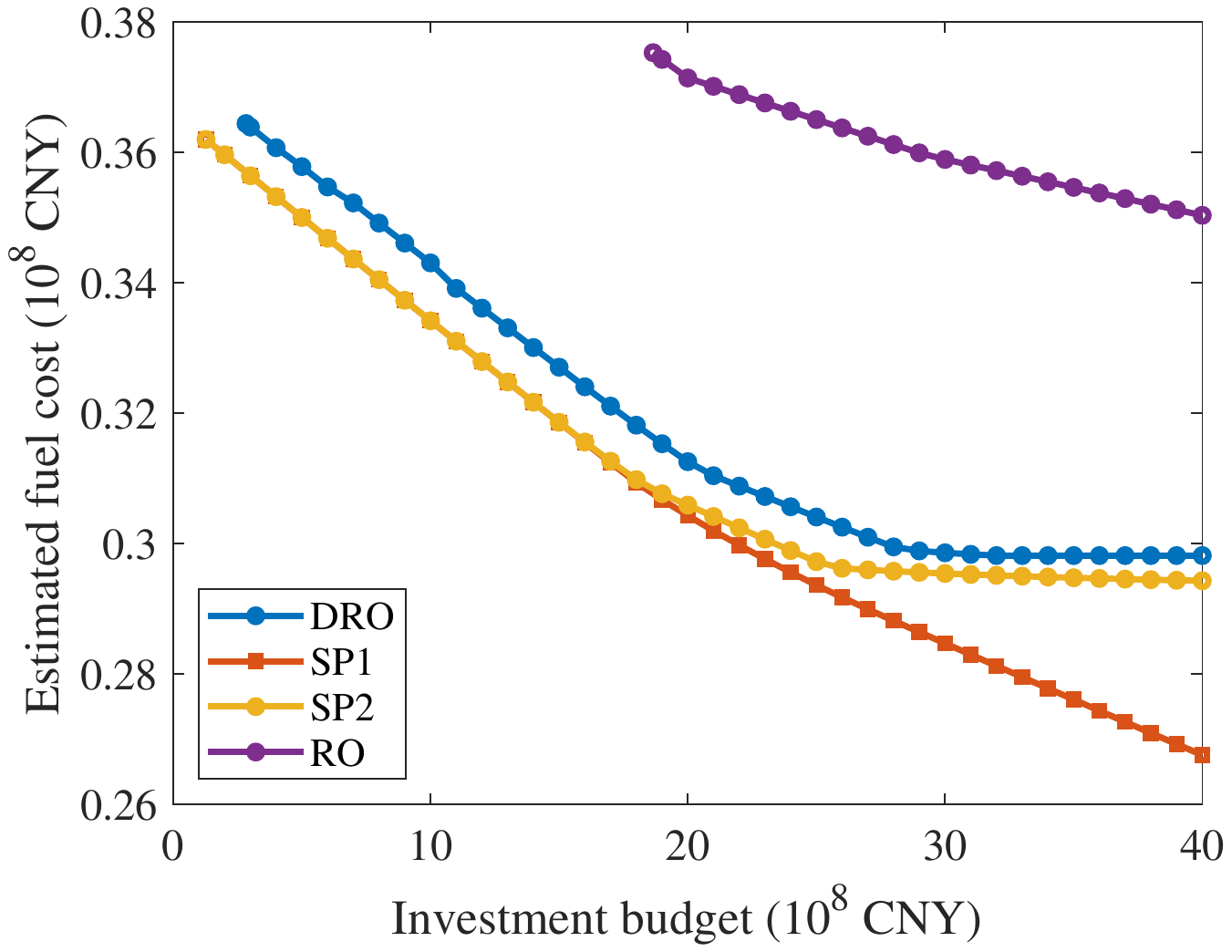}
\caption{Comparison of Pareto frontiers.}
\label{fig:Comparison}
\end{figure}

Their Pareto frontiers are depicted in Fig. \ref{fig:Comparison}. First, we compare the impact of wake effect. The Pareto frontiers of SP1 and SP2 almost coincide when the investment budget is below $17 \times 10^8$ CNY because the capacities of wind power generation are relatively small and the wake effect is not obvious, as shown in Fig. \ref{fig:Linear}. As the investment budget gets higher, the difference between the Pareto frontiers of SP1 and SP2 enlarges. The estimated fuel cost of SP1 becomes remarkably lower than that of SP2, showing that neglecting wake effect results in an over-optimistic estimate of fuel cost. Therefore, it is necessary to consider the wake effect in future power systems that will have large-scale wind power. 

Then, we compare SP2, RO, and DRO. They all take into account the wake effect but use different optimization techniques to deal with uncertainties. The Pareto frontier of DRO is above SP2 and below RO. The minimum feasible investment budget of DRO is also in-between those of SP2 and RO. This shows the results of DRO are much less conservative than RO though a little bit more costly than SP2.

In order to compare the different methods in more detail, we list the sizing results under the investment budget $30 \times 10^8$ CNY in TABLE \ref{table:comparison}, where the estimated $g^E$ and $g^N$ are obtained from the corresponding optimization methods, while the tested $g^E$ and $g^N$ are average values given by the out-of-sample tests. In terms of the estimated fuel cost $g^N$, we have SP1 $<$ SP2 $<$ DRO $<$ RO, from the most optimistic to the most conservative methods. Following this order, the total capacity of ESSs increases, and that of wind power generation decreases. This is because low-cost wind power generation brings uncertainties to the system while the ESSs can provide flexibility. The estimated load shedding $g^E$ is within the upper bound $220$ MWh under all the methods, but the tested $g^E$ of SP1 exceeds the bound. The tested $g^E$ of DRO is $73$ MWh, which is higher than that of RO but much lower than those of SP1 and SP2. 
The tested $g^N$ of SP1 and SP2 are higher than the estimated value because the distribution represented by the test data is different from the empirical distribution but the SP-based methods neglect the possible inexactness of empirical distribution. On the contrary, the tested $g^N$ under DRO is lower than the estimated one because this method already considers the inexactness of empirical distribution and the test distribution is better than the worst-case distribution. For the tested $g^N$, we have SP2 $\approx$ DRO $<$ SP1 $<$ RO. Considering that the DRO method performs better in load shedding than SP2 and has a lower cost than RO, DRO stands out among others in that it achieves a good balance between optimality and robustness.

\begin{table}[!t]
\scriptsize
\renewcommand{\arraystretch}{1.3}
\caption{Comparison of sizing results}
\label{table:comparison}
\centering
\begin{tabular}{lcccc}
\hline
Method & DRO & SP1 & SP2 & RO \\
\hline
$(x_{13}^W,x_{27}^W)$ (MW) & $(359,125)$ & $(497,41)$ & $(383,125)$ & $(99,125)$ \\
$(x_{13}^{SP},x_{23}^{SP},x_{27}^{SP})$ (MW) & $(55,0,55)$ & $(0,4,15)$ & $(40,0,10)$ & $(196,145,266)$ \\
$(x_{13}^{SE},x_{23}^{SE},x_{27}^{SE})$ (MWh) & $(119,0,73)$ & $(0,5,19)$ & $(127,0,13)$ & $(258,191,521)$ \\
Estimated $g^E$ (MWh) & $220$ & $199$ & $129$ & $0$ \\
Tested $g^E$ (MWh) & $73$ & $223$ & $133$ & $1$ \\
Estimated $g^N$ ($10^8$ CNY) & $0.2986$ & $0.2847$ & $0.2954$ & $0.3589$ \\
Tested $g^N$ ($10^8$ CNY) & $0.2959$ & $0.3045$ & $0.2957$ & $0.3326$ \\
\hline
\end{tabular}
\end{table}

\subsection{Sensitivity Analysis}

In the sensitivity analysis, we investigate the impacts of the scalar parameter $\epsilon_0$ of ambiguity set size, maximum acceptable load shedding expectation $\overline{g}^E$ in extreme conditions, and the unit cost of ESS. In this subsection, we set an upper bound $0.31 \times 10^8$ CNY for the estimated $g^N$ and minimize the investment cost. 
The results under different $\epsilon_0$ are shown in TABLE \ref{table:epsilon}. Larger $\epsilon_0$ leads to larger ambiguity sets and more conservative results, which can be seen from the increasing investment and decreasing tested $g^E$ and $g^N$. Note that the tested $g^N$ exceeds the upper bound when $\epsilon_0 = 0.01$ and $0.02$, which indicates that the ambiguity sets are not large enough to include the distribution represented by the test data. Proper parameters should be selected to establish an ambiguity set with a suitable size so that the sizing results are robust but not over-conservative.

\begin{table}[!t]
\scriptsize
\renewcommand{\arraystretch}{1.3}
\caption{Results under different $\epsilon_0$}
\label{table:epsilon}
\centering
\begin{tabular}{lcccc}
\hline
$\epsilon_0$ & $0.01$ & $0.02$ & $0.05$ & $0.08$ \\
\hline
$(x_{13}^W,x_{27}^W)$ (MW) & $(321,0)$ & $(326,0)$ & $(332,9)$ & $(332,26)$ \\
$(x_{13}^{SP},x_{23}^{SP},x_{27}^{SP})$ (MW) & $(0,0,31)$ & $(0,0,44)$ & $(13,0,74)$ & $(5,0,161)$ \\
$(x_{13}^{SE},x_{23}^{SE},x_{27}^{SE})$ (MWh) & $(0,0,41)$ & $(0,0,60)$ & $(17,0,124)$ & $(6,0,280)$ \\
Investment ($10^8$ CNY) & $18.45$ & $19.09$ & $21.26$ & $24.78$ \\
Tested $g^E$ (MWh) & $208$ & $184$ & $110$ & $37$\\
Tested $g^N$ ($10^8$ CNY) & $0.3110$ & $0.3104$ & $0.3087$ & $0.3068$ \\
\hline
\end{tabular}
\end{table}

TABLE \ref{table:shedding} shows the results under different $\overline{g}^E$. The larger the $\overline{g}^E$ is, the smaller the investment cost becomes. The capacities of ESSs change much while that of wind power generation are almost the same. The reason is that larger ESSs help to enhance reliability. Note that the tested $g^E$ is always lower than $\overline{g}^E$ and the tested $g^N$ is within the bounds, which shows the proposed method is effective.

\begin{table}[!t]
\scriptsize
\renewcommand{\arraystretch}{1.3}
\caption{Results under different $\overline{g}^E$}
\label{table:shedding}
\centering
\begin{tabular}{lcccc}
\hline
$\overline{g}^E$ (MWh) & $150$ & $220$ & $290$ & $360$ \\
\hline
$(x_{13}^W,x_{27}^W)$ (MW) & $(332,8)$ & $(332,9)$ & $(332,9)$ & $(332,9)$ \\
$(x_{13}^{SP},x_{23}^{SP},x_{27}^{SP})$ (MW) & $(7,0,147)$ & $(13,0,74)$ & $(7,0,38)$ & $(0,4,15)$ \\
$(x_{13}^{SE},x_{23}^{SE},x_{27}^{SE})$ (MWh) & $(9,0,257)$ & $(17,0,124)$ & $(9,0,49)$ & $(0,5,20)$ \\
Investment ($10^8$ CNY) & $23.44$ & $21.26$ & $19.88$ & $19.24$ \\
Tested $g^E$ (MWh) & $42$ & $110$ & $183$ & $228$ \\
Tested $g^N$ ($10^8$ CNY) & $0.3087$ & $0.3087$ & $0.3087$ & $0.3085$ \\
\hline
\end{tabular}
\end{table}

We investigate the impact of the ESS unit cost by multiplying $C^{SP}$ and $C^{SE}$ by a constant $\kappa$. The results under different $\kappa$ are listed in TABLE \ref{table:ESScost}. As $\kappa$ increases, the investment cost becomes larger, while the capacities of wind power generation and ESSs do not change much, indicating that ESSs and wind power play different roles in the system and they can hardly substitute each other even when the unit cost changes.

\begin{table}[!t]
\scriptsize
\renewcommand{\arraystretch}{1.3}
\caption{Results under different $\kappa$}
\label{table:ESScost}
\centering
\begin{tabular}{lcccc}
\hline
$\kappa$ & $0.6$ & $0.8$ & $1.0$ & $1.2$ \\
\hline
$(x_{13}^W,x_{27}^W)$ (MW) & $(332,8)$ & $(332,8)$ & $(332,9)$ & $(332,9)$ \\
$(x_{13}^{SP},x_{23}^{SP},x_{27}^{SP})$ (MW) & $(14,0,73)$ & $(13,0,74)$ & $(13,0,74)$ & $(12,0,75)$ \\
$(x_{13}^{SE},x_{23}^{SE},x_{27}^{SE})$ (MWh) & $(18,0,122)$ & $(17,0,123)$ & $(17,0,124)$ & $(16,0,125)$ \\
Investment ($10^8$ CNY) & $20.24$ & $20.75$ & $21.26$ & $21.77$ \\
\hline
\end{tabular}
\end{table}

\subsection{Scalability}

\begin{figure}[t]
\centering
\includegraphics[width=0.7\columnwidth]{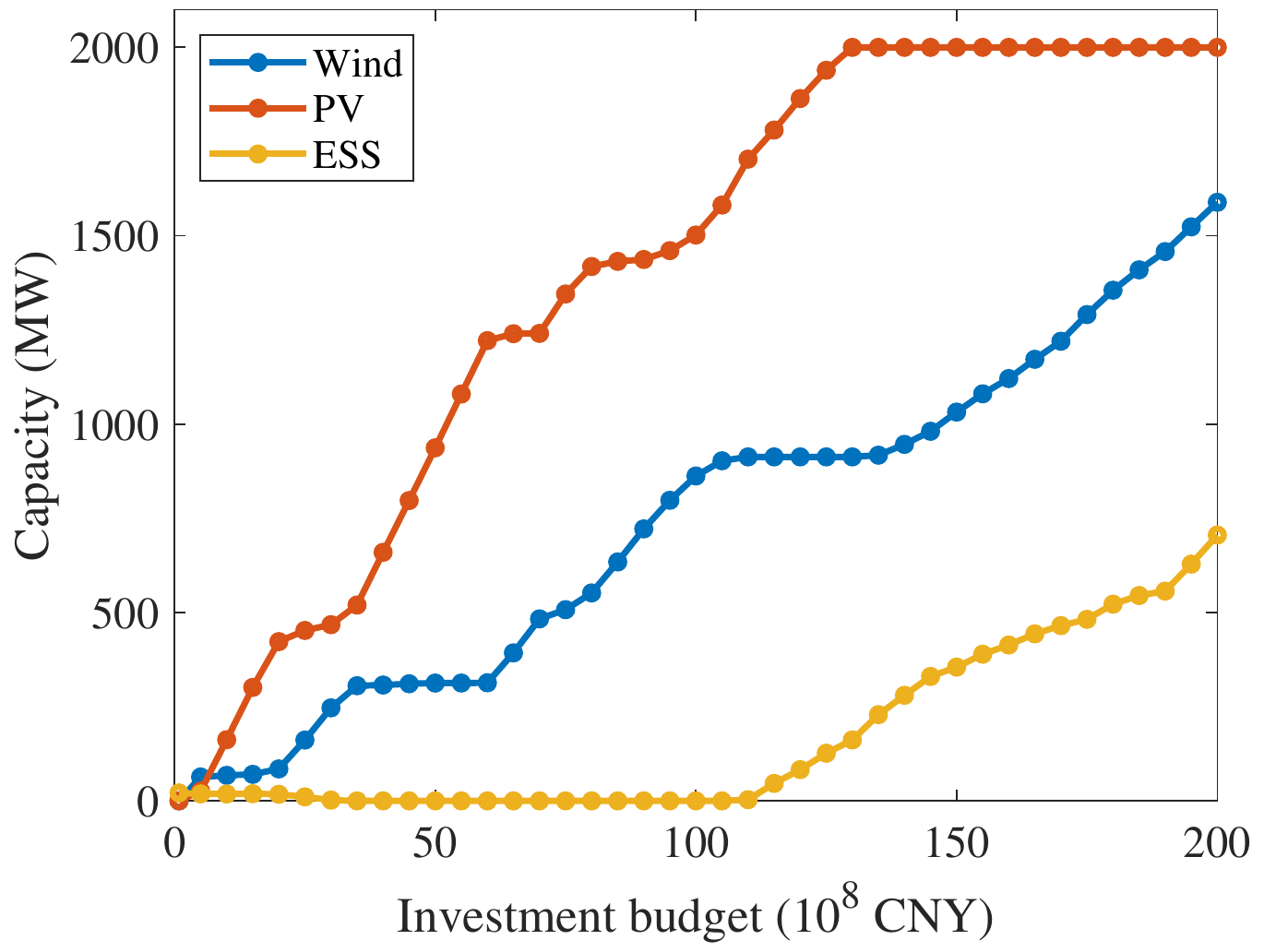}
\caption{Sizing results of the modified IEEE 118-bus system.}
\label{fig:Capacity118}
\end{figure}

To show the scalability of the proposed method, a modified IEEE 118-bus system is tested, where candidate components are $4$ wind farms, $2$ PV stations, and $3$ ESSs. The PV stations are treated in a similar way but without wake effect. The sizing results are shown in Fig. \ref{fig:Capacity118}. ESSs will be invested when the renewable capacity is large. TABLE \ref{table:scalability} shows the computation time of Algorithm \ref{alg:2} under different settings. 
Overall, the computation efficiency is acceptable for planning purposes.

\begin{table}[!t]
\scriptsize
\renewcommand{\arraystretch}{1.3}
\caption{Computation time of Algorithm \ref{alg:2} under different settings}
\label{table:scalability}
\centering
\begin{tabular}{lcccc}
\hline
Number of scenarios & $48$ & $96$ & $144$ & $192$ \\
\hline
30-bus & $31$ s & $53$ s & $145$ s & $184$ s \\
118-bus & $75$ s & $210$ s & $374$ s & $512$ s  \\
\hline
\end{tabular}
\end{table}

\section{Conclusion}
\label{sec-6}
This paper studies the sizing problem of wind power generation and energy storage in power systems considering the wake effect, which is cast as a DRO model with DDU. A solution methodology consisting of the SP approximation based on minimum Lipschitz constants, evaluations of the Lipschitz constants, and an iterative algorithm, is developed. Simulations on the modified IEEE 30-bus and 118-bus systems show the effectiveness and scalability of the proposed method, and reveal the following findings: 1) The impact of wake effect becomes more significant when the wind power capacity grows in a limited area. 2) The DRO model outperforms the traditional SP and RO models in terms of balancing optimality and robustness.
Our future work aims at considering the long-term uncertainties from society, economy, and technology.

\ifCLASSOPTIONcaptionsoff
\newpage
\fi

\bibliographystyle{IEEEtran}
\bibliography{IEEEabrv,mybib}

\clearpage
  
\appendices

\makeatletter
\@addtoreset{equation}{section}
\@addtoreset{theorem}{section}
\makeatother
\setcounter{equation}{0}  
\renewcommand{\theequation}{A.\arabic{equation}}
\renewcommand{\thetheorem}{A.\arabic{theorem}}
\section{Proof of Theorem \ref{thm:1}}
\label{appendix-A}

We omit function variable $x$ in the proof since it can be regarded as a fixed parameter of the function. We investigate the properties of the function $g^E(\zeta)$ first and then transform the distributionally robust expectation.

\subsection{Load Shedding Function}

The definition of $g^E(\zeta)$ in \eqref{eq:operation-load} can be written in a compact form as follows.
\begin{align}
    \forall \zeta \in \Xi^E, g^E(\zeta) := \min_y~ & C^\top y \nonumber \\
    \label{eq:operation-compact}
    \mbox{s.t.}~ & Ay \geq B \zeta + D,
\end{align}
where $A$, $B$, $C$, and $D$ are coefficient matrices or vectors; $y$ denotes the operation variables. Since $g^E(\zeta)$ represents the minimum load shedding, it is always nonnegative and no larger than the total load demand. Thus, the LP problem \eqref{eq:operation-compact} has a finite optimal value. Then by the strong duality of LP \cite{boyd2004convex}, we have
\begin{align}
    \forall \zeta \in \Xi^E, g^E(\zeta) = \max_\pi~ & \pi^\top B \zeta + \pi^\top D \nonumber \\
    \mbox{s.t.}~ & A^\top \pi = C, \pi \geq 0. \nonumber
\end{align}
Because the optimal value can be reached at some vertex if it is finite, 
\begin{align}
\label{eq:proof-g-vertex}
    \forall \zeta \in \Xi^E, g^E(\zeta) = \max_{\pi \in V(\Pi)}~ & \left\{ \pi^\top B \zeta + \pi^\top D \right\},
\end{align}
with polyhedron $\Pi := \{ \pi | A^\top \pi = C, \pi \geq 0 \}$ and its vertex set $V(\Pi)$. Then $V(\Pi)$ is a finite set and $g^E$ is convex \cite{boyd2004convex} and continuous. For each $\pi \in V(\Pi)$, let
\begin{align}
\Xi_\pi := \{ \zeta \in \Xi^E | g^E(\zeta) = \pi^\top B \zeta + \pi^\top D \}, \nonumber
\end{align}
then it is either empty or closed, and $\bigcup_{\pi \in V(\Pi)} \Xi_\pi = \Xi^E$.

\vspace{0.5em}

\textbf{Claim 1:} For $\pi \in V(\Pi)$, if $\Xi_\pi \neq \varnothing$, then it is convex.

Proof: Suppose $\zeta, \check{\zeta} \in \Xi_\pi$ and $\lambda \in [0,1]$. Then
\begin{align}
\label{eq:proof-convex-0}
& \pi^\top B (\lambda \zeta + (1 - \lambda) \check{\zeta}) + \pi^\top D \\
=~ & \lambda (\pi^\top B \zeta + \pi^\top D) + (1 - \lambda) (\pi^\top B \check{\zeta} + \pi^\top D) \nonumber \\
=~ & \lambda g^E(\zeta) + (1 - \lambda) g^E(\check{\zeta}) \nonumber \\
\label{eq:proof-convex-1}
\geq~ & g^E(\lambda \zeta + (1 - \lambda) \check{\zeta}) \\
\label{eq:proof-convex-2}
\geq~ & \pi^\top B (\lambda \zeta + (1 - \lambda) \check{\zeta}) + \pi^\top D,
\end{align}
where \eqref{eq:proof-convex-1} follows from the convexity of $g^E$ and \eqref{eq:proof-convex-2} is due to \eqref{eq:proof-g-vertex}. Since \eqref{eq:proof-convex-0} and \eqref{eq:proof-convex-2} are the same, the inequalities are actually equations, so $\lambda \zeta + (1 - \lambda) \check{\zeta} \in \Xi_\pi$ and therefore $\Xi_\pi$ is convex.
\qed

\vspace{0.5em}

Let $\tilde{V}$ be a subset of $V(\Pi)$ defined by
\begin{align}
\tilde{V} := \{ \pi \in V(\Pi) | \exists ~\mbox{open set}~ U ~\mbox{s.t.}~ U \subset \Xi_\pi \}. \nonumber
\end{align}
In addition, define a new function $\tilde{g}: \mathbb{R}^{M} \rightarrow \mathbb{R}$ as follows, where $M$ is the dimension of $\zeta$.
\begin{align}
\label{eq:proof-g-tilde}
\tilde{g}(\zeta) := \max_{\pi \in \tilde{V}}~ & \left\{ \pi^\top B \zeta + \pi^\top D \right\}, \forall \zeta \in \mathbb{R}^{M}.
\end{align}
We prove that $\tilde{g}(\zeta)$ is an extension of $g^E(\zeta)$.

\vspace{0.5em}

\textbf{Claim 2:} $\tilde{g}(\zeta) = g^E(\zeta), \forall \zeta \in \Xi^E$.

Proof: Denote the Lebesgue measure by $m(\cdot)$. For any $\pi \in V(\Pi)$, if $m(\Xi_\pi) > 0$, then $\Xi_\pi$ contains some open set since it is convex. Additionally, $|V(\Pi)| < \infty$, so $m(\bigcup_{\pi \in V(\Pi) \setminus \tilde{V}} \Xi_\pi) = 0$. Thus, $m(\bigcup_{\pi \in \tilde{V}} \Xi_\pi) = m(\bigcup_{\pi \in V(\Pi)} \Xi_\pi) = m(\Xi^E) > 0$ and it is finite since $\Xi^E$ is bounded. Note that $\Xi^E$ is convex, $\tilde{g} = g^E$ on $\bigcup_{\pi \in \tilde{V}} \Xi_\pi$, and they are continuous, so $\tilde{g}(\zeta) = g^E(\zeta), \forall \zeta \in \Xi^E$.
\qed

\vspace{0.5em}

Because $\tilde{g}(\zeta)$ is proper, convex, and continuous, according to the property of conjugate function \cite{bertsekas2009convex}, $\tilde{g} = \tilde{g}^{**}$, which is the bi-conjugate function, i.e.,
\begin{align}
    \tilde{g}(\zeta) = \sup_{\theta \in \Theta} \{ \theta^\top \zeta - \tilde{g}^*(\theta) \}, \forall \zeta \in \mathbb{R}^{M}, \nonumber
\end{align}
where the conjugate function $\tilde{g}^*(\theta) := \sup_{\zeta \in \mathbb{R}^{M}} \{ \theta^\top \zeta - \tilde{g}(\zeta) \}$ and $\Theta := \{ \theta | \tilde{g}^*(\theta) < \infty \}$. 

\vspace{0.5em}

\textbf{Claim 3:} $\Theta = \mbox{conv}\{ B^\top \pi | \pi \in \tilde{V} \}$, i.e., the convex hull of $\{ B^\top \pi | \pi \in \tilde{V} \}$.

Proof: (1) Prove that $\mbox{conv}\{ B^\top \pi | \pi \in \tilde{V}\} \subset \Theta$. 

For any $\theta \in \mbox{conv}\{ B^\top \pi | \pi \in \tilde{V}\}$, there exist $\pi_i \in \tilde{V}$ and $\lambda_i, i \in \mathcal{I}$ subject to $\lambda_i \geq 0, i \in \mathcal{I}$, $\sum_{i \in \mathcal{I}} \lambda_i = 1$ and $\theta = \sum_{i \in \mathcal{I}} \lambda_i B^\top \pi_i$, so
\begin{align}
\tilde{g}(\zeta) =~ & \max_{\pi \in \tilde{V}} \{ \pi^\top B \zeta + \pi^\top D \} \nonumber \\
=~ & \sum_{i \in \mathcal{I}} \lambda_i \max_{\pi \in \tilde{V}} \{ \pi^\top B \zeta + \pi^\top D \} \nonumber \\
\geq~ & \sum_{i \in \mathcal{I}} \lambda_i (\pi_i^\top B \zeta + \pi_i^\top D) \nonumber \\
=~ & \theta^\top \zeta + \sum_{i \in \mathcal{I}} \lambda_i \pi_i^\top D, \nonumber
\end{align}
and then
\begin{align}
\sup_{\zeta \in \mathbb{R}^{M}} \{ \theta^\top \zeta - \tilde{g}(\zeta) \} \leq - \sum_{i \in \mathcal{I}} \lambda_i \pi_i^\top D \leq \max_{\pi \in \tilde{V}} |\pi^\top D| < + \infty, \nonumber
\end{align}
which means $\theta \in \Theta$. Thus, $\mbox{conv}\{ B^\top \pi | \pi \in \tilde{V}\} \subset \Theta$.

(2) Prove that $\Theta \subset \mbox{conv}\{ B^\top \pi | \pi \in \tilde{V}\}$.

Assume $\theta \notin \mbox{conv}\{ B^\top \pi | \pi \in \tilde{V}\}$. Since $\tilde{V}$ is a finite set, $\mbox{conv}\{ B^\top \pi | \pi \in \tilde{V}\}$ is closed and convex. Then $\theta$ and $\mbox{conv}\{ B^\top \pi | \pi \in \tilde{V}\}$ can be strictly separated \cite{boyd2004convex}, i.e., there are vector $a \neq 0$ and constant $\epsilon > 0$ such that $\theta^\top a - \epsilon \geq \pi^\top B a, \forall \pi \in \tilde{V}$. Then for any $\lambda > 0$, we have
\begin{align}
\tilde{g}(\lambda a) ~=~ & \max_{\pi \in \tilde{V}} \{ \pi^\top B (\lambda a) + \pi^\top D \} \nonumber \\
\leq~ & \max_{\pi \in \tilde{V}} \{ \lambda(\theta^\top a - \epsilon) + \pi^\top D \} \nonumber \\
=~ & \lambda(\theta^\top a - \epsilon) + \max_{\pi \in \tilde{V}} \{ \pi^\top D \}, \nonumber
\end{align}
then
\begin{align}
\theta^\top (\lambda a) - \tilde{g}(\lambda a)
\geq~ & \theta^\top (\lambda a) - \lambda(\theta^\top a - \epsilon) - \max_{\pi \in \tilde{V}} \{ \pi^\top D \} \nonumber \\
=~ & \lambda \epsilon - \max_{\pi \in \tilde{V}} \{ \pi^\top D \} \rightarrow + \infty~ (\lambda \rightarrow + \infty), \nonumber
\end{align}
which means $\theta \notin \Theta$. Thus, $\Theta \subset \mbox{conv}\{ B^\top \pi | \pi \in \tilde{V}\}$.
\qed

\vspace{0.5em}

\textbf{Claim 4:} $L_i^{WE} = \sup_{\theta \in \Theta} \| \theta_i^W \|_\infty = \max_{\pi \in \tilde{V}} \| (B^\top \pi)_i^W \|_\infty$, where $(B^\top \pi)_i^W$ denotes the projection of $(B^\top \pi)$ to the component corresponding to the wind farm at bus $i$.

Proof: $\sup_{\theta \in \Theta} \| \theta_i^W \|_\infty = \max_{\pi \in \tilde{V}} \| (B^\top \pi)_i^W \|_\infty$ directly follows from $\Theta = \mbox{conv}\{ B^\top \pi | \pi \in \tilde{V} \}$ and that $\tilde{V}$ is finite. We prove $L_i^{WE} = \max_{\pi \in \tilde{V}} \| (B^\top \pi)_i^W \|_\infty$.

(1) Prove that $L_i^{WE} \geq \max_{\pi \in \tilde{V}} \| (B^\top \pi)_i^W \|_\infty$.

There is a $\pi_0 \in \tilde{V}$ such that
\begin{align}
\| (B^\top \pi_0)_i^W \|_\infty = \max_{\pi \in \tilde{V}} \| (B^\top \pi)_i^W \|_\infty. \nonumber
\end{align}

According to the definition of $\tilde{V}$, there is an open set $U \subset \Xi_{\pi_0} \subset \Xi^E$. For $\zeta, \check{\zeta} \in U$, $g^E(\zeta) = \tilde{g}(\zeta) = \pi_0^\top B \zeta + \pi_0^\top D$ and $g^E(\check{\zeta}) = \tilde{g}(\check{\zeta}) = \pi_0^\top B \check{\zeta} + \pi_0^\top D$. Since $U$ is open, we can choose $\zeta$ and $\check{\zeta}$ so that $\zeta^D = \check{\zeta}^D$, $\zeta_j^W = \check{\zeta}_j^W$ for any $j \neq i$ and
\begin{align}
|g^E(\zeta) - g^E(\check{\zeta})| & = |\pi_0^\top B (\zeta - \check{\zeta})| \nonumber \\
& = \| (B^\top \pi_0)_i^W \|_\infty \| (\zeta - \check{\zeta})_i^W \|_1 \nonumber \\
& = \| (B^\top \pi_0)_i^W \|_\infty \| \zeta - \check{\zeta} \|_1, \nonumber
\end{align}
due to the property of dual norm. Then by the definition of $L_i^{WE}$, we have $L_i^{WE} \geq \| (B^\top \pi_0)_i^W \|_\infty = \max_{\pi \in \tilde{V}} \| (B^\top \pi)_i^W \|_\infty$.

(2) Prove that $L_i^{WE} \leq \max_{\pi \in \tilde{V}} \| (B^\top \pi)_i^W \|_\infty$.

For arbitrary $\zeta, \check{\zeta} \in \Xi^E$ subject to $\zeta^D = \check{\zeta}^D$ and $\zeta_j^W = \check{\zeta}_j^W$ for any $j \neq i$, there are $\pi_0$ and $\check{\pi}_0$ so that
\begin{align}
& g^E(\zeta) = \tilde{g}(\zeta) = \pi_0^\top B \zeta + \pi_0^\top D \geq \check{\pi}_0^\top B \zeta + \check{\pi}_0^\top D, \nonumber \\
& g^E(\check{\zeta}) = \tilde{g}(\check{\zeta}) = \check{\pi}_0^\top B \check{\zeta} + \check{\pi}_0^\top D \geq \pi_0^\top B \check{\zeta} + \pi_0^\top D, \nonumber
\end{align}
so
\begin{align}
\check{\pi}_0^\top B (\zeta - \check{\zeta}) \leq g^E(\zeta) - g^E(\check{\zeta}) \leq \pi_0^\top B (\zeta - \check{\zeta}), \nonumber
\end{align}
and then
\begin{align}
|g^E(\zeta) - g^E(\check{\zeta})| \leq~ & \max_{\pi \in \tilde{V}} |\pi^\top B (\zeta - \check{\zeta})| \nonumber \\
\leq~ & \max_{\pi \in \tilde{V}} \| (B^\top \pi)_i^W \|_\infty \cdot \|\zeta - \check{\zeta} \|_1. \nonumber
\end{align}
Since $\zeta$ and $\check{\zeta}$ are arbitrary, we have $L_i^{WE} \leq \max_{\pi \in \tilde{V}} \| (B^\top \pi)_i^W \|_\infty$.
\qed

\vspace{0.5em}

Similarly, $L^{DE} = \sup_{\theta \in \Theta} \| \theta^D \|_\infty = \max_{\pi \in \tilde{V}} \| (B^\top \pi)^D \|_\infty$.

Remark: The fuel cost function $g^N$ can be handled in a similar way when the dimension of $\Xi^N$ is $M$. 

\subsection{Distributionally Robust Expectation}

The proof in this subsection is analogous to that in \cite{esfahani2018data}.

By the definition of $\mathcal{B}^E(\epsilon^E)$ in \eqref{eq:ambiguity}, the distributionally robust expectation is
\begin{align}
    & \sup_{\mathbb{P} \in \mathcal{B}^E (\epsilon^E)} \mathbb{E}_\mathbb{P} [g^E(\zeta)] = \sup_{\mathbb{P} \in \mathcal{B}^E (\epsilon^E)} \mathbb{E}_\mathbb{P} [\tilde{g}(\zeta)] \nonumber \\
    =~ & \left\{
    \begin{aligned}
    \sup~ & \int_{\Xi_i^{WE} \times \Xi^D} \tilde{g}(\zeta) \left( \frac{1}{|S^E|} \sum_{n \in S^E} \left( \bigotimes_{i \in S^B}  \mathbb{P}_{i,n}^W \right) \otimes \mathbb{P}_n^D \right) (d\zeta) \\
    \mbox{s.t.}~ & \mathbb{P}_{i,n}^W \in \mathcal{M} (\Xi_i^{WE}), \Pi_i^W = \frac{1}{|S^E|} \sum_{n \in S^E} \mathbb{P}_{i,n}^W \otimes \delta_{\hat{\zeta}_{i,n}^W} \\
    & \int_{(\Xi_i^{WE})^2} \|\zeta_i^W - \tilde{\zeta}_i^W \|_1 \Pi_i^W(d\zeta_i^W, d\tilde{\zeta}_i^W) \leq \epsilon_i^{WE} x_i^W \\
    & \mathbb{P}_n^D \in \mathcal{M}(\Xi^{DE}), \Pi^D = \frac{1}{|S^E|} \sum_{n \in S^E} \mathbb{P}_n^D \otimes \delta_{\hat{\zeta}_n^D} \\
    & \int_{(\Xi^{DE})^2} \| \zeta^D - \tilde{\zeta}^D \|_1 \Pi^D (d\zeta^D, d\tilde{\zeta}^D) \leq \epsilon^{DE} \\
    \end{aligned}
    \right. \nonumber \\
    \label{eq:proof-2}
    =~ & \left\{
    \begin{aligned}
    \sup~ & \frac{1}{|S^E|} \sum_{n \in S^E} \int_{\Xi^E} \tilde{g}(\zeta) \mathbb{P}_n (d\zeta) \\
    \mbox{s.t.}~ & \mathbb{P}_{i,n}^W \in \mathcal{M} (\Xi_i^{WE}), \mathbb{P}_n^D \in \mathcal{M}(\Xi^{DE}) \\
    & \mathbb{P}_n = \left( \bigotimes_{i \in S^B}  \mathbb{P}_{i,n}^W \right) \otimes \mathbb{P}_n^D \\
    & \frac{1}{|S^E|} \sum_{n \in S^E} \int_{\Xi_i^{WE}} \|\zeta_i^W - \hat{\zeta}_{i,n}^W \|_1 \mathbb{P}_{i,n}^W(d\zeta_i^W) \leq \epsilon_i^{WE} x_i^W \\
    & \frac{1}{|S^E|} \sum_{n \in S^E} \int_{\Xi^{DE}} \| \zeta^D - \hat{\zeta}_n^D \|_1 \mathbb{P}_n^D (d\zeta^D) \leq \epsilon^{DE} \\
    \end{aligned}
    \right.,
\end{align}
where $\Xi^E = \left( \times_{i \in S^B} \Xi_i^{WE} \right) \times \Xi^{DE}$; \eqref{eq:proof-2} follows from the linearity of expectation and the definitions of Dirac distribution and joint distribution.

Using the duality technique, we have
\begin{align}
    & \sup_{\mathbb{P} \in \mathcal{B}^E (\epsilon^E)} \mathbb{E}_\mathbb{P} [g^E(\zeta)] \nonumber \\
    \label{eq:proof-3}
    =~ & \left\{
    \begin{aligned}
    \sup_{\mathbb{P}} \inf_{\lambda}~ & \frac{1}{|S^E|} \sum_{n \in S^E} \int_{\Xi^E} \tilde{g}(\zeta) \mathbb{P}_n (d\zeta) + \sum_{i \in S^B} \lambda_i^W \{ \epsilon_i^{WE} x_i^W \\
    & - \frac{1}{|S^E|} \sum_{n \in S^E} \int_{\Xi_i^{WE}} \|\zeta_i^W - \hat{\zeta}_{i,n}^W \|_1 \mathbb{P}_{i,n}^W(d\zeta_i^W) \} \\
    & + \lambda^D \{ \epsilon^{DE} - \frac{1}{|S^E|} \sum_{n \in S^E} \int_{\Xi^{DE}} \| \zeta^D - \hat{\zeta}_n^D \|_1 \mathbb{P}_n^D (d\zeta^D) \} \\
    \mbox{s.t.}~ & \mathbb{P}_{i,n}^W \in \mathcal{M} (\Xi_i^{WE}), \mathbb{P}_n^D \in \mathcal{M}(\Xi^{DE}), \lambda_i^W \geq 0, \lambda^D \geq 0 \\
    & \mathbb{P}_n = \left( \bigotimes_{i \in S^B}  \mathbb{P}_{i,n}^W \right) \otimes \mathbb{P}_n^D \\
    \end{aligned}
    \right. \\
    \label{eq:proof-4}
    =~ & \left\{
    \begin{aligned}
    \inf_{\lambda} \sup_{\mathbb{P}}~ & \sum_{i \in S^B} \lambda_i^W \epsilon_i^{WE} x_i^W + \lambda^D \epsilon^{DE} \\
    & + \frac{1}{|S^E|} \sum_{n \in S^E} \left( \int_{\Xi^E} \tilde{g}(\zeta) \mathbb{P}_n (d\zeta) \right. \\
    & - \sum_{i \in S^B} \lambda_i^W \int_{\Xi_i^{WE}} \|\zeta_i^W - \hat{\zeta}_{i,n}^W \|_1 \mathbb{P}_{i,n}^W(d\zeta_i^W) \\
    & \left. - \lambda^D \int_{\Xi^{DE}} \| \zeta^D - \hat{\zeta}_n^D \|_1 \mathbb{P}_n^D (d\zeta^D) \right) \\
    \mbox{s.t.}~ & \mathbb{P}_{i,n}^W \in \mathcal{M} (\Xi_i^{WE}), \mathbb{P}_n^D \in \mathcal{M}(\Xi^{DE}), \lambda_i^W \geq 0, \lambda^D \geq 0 \\
    & \mathbb{P}_n = \left( \bigotimes_{i \in S^B}  \mathbb{P}_{i,n}^W \right) \otimes \mathbb{P}_n^D \\
    \end{aligned}
    \right. \\
    \label{eq:proof-5}
    =~ & \left\{
    \begin{aligned}
    \inf_{\lambda}~ & \sum_{i \in S^B} \lambda_i^W \epsilon_i^{WE} x_i^W + \lambda^D \epsilon^{DE} \\
    & + \frac{1}{|S^E|} \sum_{n \in S^E} \sup_{\zeta \in \Xi^E} \{ \tilde{g}(\zeta) - \sum_{i \in S^B} \lambda_i^W \|\zeta_i^W - \hat{\zeta}_{i,n}^W \|_1 \\
    & - \lambda^D \| \zeta^D - \hat{\zeta}_n^D \|_1 \} \\
    \mbox{s.t.}~ & \lambda_i^W \geq 0, \lambda^D \geq 0 \\
    \end{aligned}
    \right..
\end{align}
From \eqref{eq:proof-3} to \eqref{eq:proof-4}, the ``$\leq$" direction uses the max-min inequality, and the opposite direction follows from the strong duality result of moment problems \cite{shapiro2001on}. The Dirac distribution $\delta_\zeta$ for $\zeta \in \Xi^E$ can be regarded as a product of Dirac distributions on $\Xi_i^{WE}$ and $\Xi^{DE}$, so \eqref{eq:proof-5} holds.

Using $\tilde{g} = \tilde{g}^{**}$, we have
\begin{align}
    & \sup_{\zeta \in \Xi^E} \{ \tilde{g}(\zeta) - \sum_{i \in S^B} \lambda_i^W \|\zeta_i^W - \hat{\zeta}_{i,n}^W \|_1 - \lambda^D \| \zeta^D - \hat{\zeta}_n^D \|_1 \} \nonumber \\
    =~ & \sup_{\zeta \in \Xi^E}~ \sup_{\theta \in \Theta} \{ \theta^\top \zeta - \tilde{g}^*(\theta) - \sum_{i \in S^B} \lambda_i^W \|\zeta_i^W - \hat{\zeta}_{i,n}^W \|_1 \nonumber \\
    & - \lambda^D \| \zeta^D - \hat{\zeta}_n^D \|_1 \} \nonumber \\
    =~ & \sup_{\zeta \in \Xi^E}~ \sup_{\theta \in \Theta}~ \inf_{\mu: \| \mu_i^W \|_\infty \leq \lambda_i^W, \| \mu^D \|_\infty \leq \lambda^D} \{ \theta^\top \zeta - \tilde{g}^*(\theta) \nonumber \\
    \label{eq:proof-12}
    & + \sum_{i \in S^B} ( (\mu_i^W)^\top \zeta_i^W - (\mu_i^W)^\top \hat{\zeta}_{i,n}^W) + (\mu^D)^\top \zeta^D - (\mu^D)^\top \hat{\zeta}_n^D \} \\
    =~ & \sup_{\zeta \in \Xi^E}~ \sup_{\theta \in \Theta}~ \inf_{\mu: \genfrac{}{}{0pt}{}{\| \mu_i^W \|_\infty \leq \lambda_i^W}{\| \mu^D \|_\infty \leq \lambda^D}} \{ (\theta + \mu)^\top \zeta - \tilde{g}^*(\theta) - \mu^\top \hat{\zeta}_n \} \nonumber \\
    \label{eq:proof-14}
    =~ & \sup_{\theta \in \Theta}~ \inf_{\mu: \genfrac{}{}{0pt}{}{\| \mu_i^W \|_\infty \leq \lambda_i^W}{\| \mu^D \|_\infty \leq \lambda^D}}~ \sup_{\zeta \in \Xi^E} \{ (\theta + \mu)^\top \zeta - \tilde{g}^*(\theta) - \mu^\top \hat{\zeta}_n \} \\
    \label{eq:proof-whole}
    \leq~ & \sup_{\theta \in \Theta}~ \inf_{\mu: \genfrac{}{}{0pt}{}{\| \mu_i^W \|_\infty \leq \lambda_i^W}{\| \mu^D \|_\infty \leq \lambda^D}}~  \sup_{\zeta \in \mathbb{R}^{M}}  \{(\theta + \mu)^\top \zeta  - \tilde{g}^*(\theta) - \mu^\top \hat{\zeta}_n \},
\end{align}
where \eqref{eq:proof-12} follows from the property of dual norm; \eqref{eq:proof-14} is a consequence of the minimax theorem \cite{bertsekas2009convex}. Since
\begin{align}
\sup_{\zeta \in \mathbb{R}^{M}} \{ (\theta + \mu)^\top \zeta \} = \left\{
\begin{aligned}
& 0, ~\mbox{if}~ \mu = - \theta, \\
& +\infty, ~\mbox{otherwise},
\end{aligned}
\right. \nonumber
\end{align}
we have \eqref{eq:proof-whole} $ = \sup_{\theta \in \Theta} \{ - \tilde{g}^*(\theta) + \theta^\top \hat{\zeta}_n \} = \tilde{g}(\hat{\zeta}_n) = g^E(\hat{\zeta}_n)$ if $\| \theta_i^W \|_\infty \leq \lambda_i^W$ and $ \| \theta^D \|_\infty \leq \lambda^D$, $ \forall \theta \in \Theta$; otherwise, \eqref{eq:proof-whole} $ = + \infty$. Therefore,
\begin{align}
    & \sup_{\mathbb{P} \in \mathcal{B}^E (\epsilon^E)} \mathbb{E}_\mathbb{P} [g^E(\zeta)] \nonumber \\
    \leq~ & \left\{
    \begin{aligned}
    \inf_{\lambda}~ & \sum_{i \in S^B} \lambda_i^W \epsilon_i^{WE} x_i^W + \lambda^D \epsilon^{DE} \\
    & + \frac{1}{|S^E|} \sum_{n \in S^E} \sup_{\zeta \in \mathbb{R}^{M}} \{ \tilde{g}(\zeta) - \sum_{i \in S^B} \lambda_i^W \|\zeta_i^W - \hat{\zeta}_{i,n}^W \|_1 \\
    & - \lambda^D \| \zeta^D - \hat{\zeta}_n^D \|_1 \} \\
    \mbox{s.t.}~ & \lambda_i^W \geq 0, \lambda^D \geq 0 \\
    \end{aligned}
    \right. \nonumber \\
    =~ & \sum_{i \in S^B} \left( \sup_{\theta \in \Theta}  \| \theta_i^W \|_\infty \right) \epsilon_i^{WE} x_i^W \nonumber \\
    & + \left( \sup_{\theta \in \Theta}  \| \theta^D \|_\infty \right) \epsilon^{DE} + \frac{1}{|S^E|} \sum_{n \in S^E} g^E(\hat{\zeta}_n) \nonumber \\
    =~ & \sum_{i \in S^B} L_i^{WE} \epsilon_i^{WE} x_i^W + L^{DE} \epsilon^{DE} + \mathbb{E}_{\mathbb{P}_0^E}[g^E(\zeta)], \nonumber
\end{align}
by Claim 4. This completes the proof.

\makeatletter
\@addtoreset{equation}{section}
\@addtoreset{theorem}{section}
\makeatother
\setcounter{equation}{0}  
\renewcommand{\theequation}{B.\arabic{equation}}
\renewcommand{\thetheorem}{B.\arabic{theorem}}
\section{Calculating Lipschitz Constants}
\label{appendix-B}

The methods of calculating Lipschitz constants are different for $g^E$ and $g^N$ since their domains are defined in different ways.

\subsection{Lipschitz Constants of the Load Shedding Function}

The definition of load shedding function in \eqref{eq:operation-load} can be written in the following compact form:
\bsq
\label{eq:L-primal}
\begin{align}
g^E(x, \zeta) := \min_y~ & C^\top y \\
\mbox{s.t.}~ & A_1 y \geq B_1 x + B_2 \zeta + B_3 \\
& A_2 y = B_4 x + B_5 \zeta + B_6,
\end{align}
\esq
where $A_1$, $A_2$, $B_1$, $B_2$, $B_3$, $B_4$, $B_5$, $B_6$, and $C$ are coefficient matrices or vectors; $y$ represents the operation variables. The dual problem of \eqref{eq:L-primal} is
\bsq
\label{eq:L-dual}
\begin{align}
\max_{\lambda,\mu}~ & (B_1 x + B_2 \zeta + B_3)^\top \lambda + (B_4 x + B_5 \zeta + B_6)^\top \mu \\
\mbox{s.t.}~ & A_1^\top \lambda + A_2^\top \mu = C, \lambda \geq 0.
\end{align}
\esq

Because the LP problem \eqref{eq:L-primal} has a finite optimal value for any $x \in \mathcal{X}$ and $\zeta \in \Xi^E$, the strong duality holds \cite{boyd2004convex}. Then for dual optimal solution $(\lambda, \mu)$, we have
\begin{align}
g^E(x, \zeta) =~ & (B_1 x + B_2 \zeta + B_3)^\top \lambda + (B_4 x + B_5 \zeta + B_6)^\top \mu \nonumber \\
=~ & (B_2^\top \lambda + B_5^\top \mu)^\top \zeta + (B_1 x + B_3)^\top \lambda + (B_4 x + B_6)^\top \mu, \nonumber
\end{align}
which shows that the value of $B_2^\top \lambda + B_5^\top \mu$ reflects how $g^E(x, \zeta)$ changes as $\zeta$ varies.

For fixed $x \in \mathcal{X}$, in order to find a Lipschitz constant $\tilde{L}_i^{WE}(x)$, we solve the following optimization problem.
\bsq
\label{eq:L-max}
\begin{align}
\label{eq:L-max-1}
\max_{\zeta, y, \lambda, \mu, \gamma}~ & - \gamma_a \\
\label{eq:L-max-2}
\mbox{s.t.}~ & \zeta \in \Xi^E, \gamma = B_2^\top \lambda + B_5^\top \mu \\
\label{eq:L-max-3}
& 0 \leq A_1 y - B_1 x - B_2 \zeta - B_3 \perp \lambda \geq 0 \\
\label{eq:L-max-4}
& A_2 y = B_4 x + B_5 \zeta + B_6, A_1^\top \lambda + A_2^\top \mu = C,
\end{align}
\esq
where the decision variables are $\zeta$, $y$, $\lambda$, $\mu$, and $\gamma$. In \eqref{eq:L-max-1}, $\gamma_a$ represents a component corresponding to $\zeta_i^W$. Any such component works because all $t \in \{1, 2, \cdots, T\}$ lead to equivalent problems, which can be observed from \eqref{eq:operation-load} and the power system operation constraints. Since larger wind power output results in smaller load shedding, $- \gamma_a$ is maximized in \eqref{eq:L-max-1}. Constraints \eqref{eq:L-max-3} and \eqref{eq:L-max-4} form the KKT condition of problem \eqref{eq:L-primal}, where \eqref{eq:L-max-3} can be linearized by the big-M technique \cite{pineda2019solving}, so that \eqref{eq:L-max} is transformed into an MILP problem. It can also be solved directly as a bilinear program using solvers. Its optimal value or any upper bound can be the desired $\tilde{L}_i^{WE}(x)$.

We can also find a uniform bound $\tilde{L}_i^{WE}$ so that $\tilde{L}_i^{WE} \geq \tilde{L}_i^{WE}(x), \forall x \in \mathcal{X}$ by the optimization problem below.
\bsq
\begin{align}
\max_{x, \zeta, y, \lambda, \mu, \gamma}~ & - \gamma_a \\
\mbox{s.t.}~ & x \in \mathcal{X}, \eqref{eq:L-max-2}-\eqref{eq:L-max-4},
\end{align}
\esq
which is modified from \eqref{eq:L-max} by allowing $x$ to vary within $\mathcal{X}$. This problem can still be either directly solved as a bilinear program or transformed into an MILP problem and then solved.

\subsection{Lipschitz Constants of the Fuel Cost Function}

The definition of the fuel cost function $g^N(x, \zeta)$ can also be written in the form of \eqref{eq:L-primal}, and so is the dual problem \eqref{eq:L-dual}. When $x$ is fixed, $g^N(x, \zeta)$ is a convex and piecewise linear function of $\zeta$. According to the relation of Lipschitz continuity and local Lipschitz continuity \cite{nguyen2021lipschitz}, the maximum local Lipschitz constant at $\hat{\zeta}_n, n \in S^N$ is also a Lipschitz constant on $\Xi^N$. Therefore, solve the dual problem \eqref{eq:L-dual} at each $\hat{\zeta}_n, n \in S^N$ to find the dual optimal solutions $(\hat{\lambda}_n(x), \hat{\mu}_n(x)), n \in S^N$, where $x$ emphasizes that they are dependent on $x$. Then calculate $- (B_2^\top \hat{\lambda}_n(x) + B_5^\top \hat{\mu}_n(x) )$, extract the component corresponding to $\zeta_i^W$, and denote it by $\hat{\gamma}_{i,n}^W(x)$. Note that $\hat{\gamma}_{i,n}^W(x) = ( \hat{\gamma}_{i,t,n}^W(x); t \in S^T )$ is a vector with $T$ dimensions. Then $\tilde{L}_i^{WN}(x)$ can be estimated by $\max_{n \in S^N} \max_{t \in S^T} \hat{\gamma}_{i,t,n}^W(x)$.
The Lipschitz constant $\tilde{L}^{DN}(x)$ can be estimated in a similar way.

\end{document}